\documentclass[]{article}

\usepackage{amsmath, amssymb, dsfont, mathrsfs,amsthm}
\usepackage{float}
\usepackage{mdframed}
\usepackage{mathtools}
\usepackage{appendix}
\usepackage[hidelinks]{hyperref}
\usepackage{graphicx}
\usepackage{caption}
\usepackage[T1]{fontenc}
\usepackage[margin=1in]{geometry}
\newcommand{\mnoteDone}[1]{}
\newcommand{\jnoteDone}[1]{}
\newcommand{\mfnoteDone}[1]{}
\newcommand{\jfnoteDone}[1]{}
\newcommand{\mTODODone}[1]{}
\newcommand{\jTODODone}[1]{}

\allowdisplaybreaks

\title{An Overview of Capacity Results for Synchronization Channels
}

\author{Mahdi Cheraghchi\thanks{EECS Department, University of Michigan. Email: mahdich@umich.edu} \and Jo\~ao Ribeiro\thanks{Department of Computing, Imperial College London. Email: j.lourenco-ribeiro17@imperial.ac.uk}}

\date{}

\newcommand{\cC}{\mathcal{C}}

\newcommand{\cX}{\mathcal{X}}
\newcommand{\cY}{\mathcal{Y}}

\newcommand{\cS}{\mathcal{S}}

\newcommand{\cU}{\mathcal{U}}

\newcommand{\E}{\mathds{E}}
\newcommand{\supp}{\mathsf{supp}}

\newcommand{\eps}{\epsilon}

\newcommand{\cP}{\mathcal{P}}

\newcommand{\cV}{\mathcal{V}}

\newcommand{\Dec}{\mathsf{Dec}}

\newcommand{\cA}{\mathcal{A}}
\newcommand{\LCS}{\mathsf{LCS}}

\newtheorem{thm}{Theorem}

\newtheorem{lem}[thm]{Lemma}

\newtheorem{coro}[thm]{Corollary}

\newtheorem{defn}[thm]{Definition}
\newtheorem{remark}[thm]{Remark}

\newcommand{\Ch}{\mathsf{Ch}}
\newcommand{\Ca}{\mathsf{Cap}}
\newcommand{\NB}{\mathsf{NB}}
\newcommand{\Bin}{\mathsf{Bin}}
\newcommand{\Geom}{\mathsf{Geom}}
\newcommand{\Ber}{\mathsf{Ber}}

\newcommand{\KL}{D_\mathsf{KL}}
\newcommand{\bits}{\{0,1\}}
\newcommand{\Del}{\mathsf{Del}}
\newcommand{\Poi}{\mathsf{Poi}}
\newcommand{\InvBin}{\mathsf{InvBin}}

\let\originalleft\left
\let\originalright\right
\renewcommand{\left}{\mathopen{}\mathclose\bgroup\originalleft}
\renewcommand{\right}{\aftergroup\egroup\originalright}

\usepackage{url}
\begin{document}

\maketitle

\begin{abstract}
Synchronization channels, such as the well-known deletion channel, are surprisingly harder to analyze than memoryless channels, and they are a source of many fundamental problems in information theory and theoretical computer science.

One of the most basic open problems regarding synchronization channels is the derivation of an exact expression for their capacity.
Unfortunately, most of the classic information-theoretic techniques at our disposal fail spectacularly when applied to synchronization channels.
Therefore, new approaches must be considered to tackle this problem.
This survey gives an account of the great effort made over the past few decades to better understand the (broadly defined) capacity of synchronization channels, including both the main results and the novel techniques underlying them.
Besides the usual notion of channel capacity, we also discuss the zero-error capacity of synchronization channels.
\end{abstract}


\section{Introduction}\label{sec:intro}

Synchronization channels are communication channels that induce a loss of synchronization between sender and receiver.
In general, this means that there is memory between different outputs of the channel, even if the inputs to the channel are independent and identically distributed (i.i.d.).
As a result, synchronization channels appear much more difficult to analyze than memoryless channels like the Binary Symmetric Channel (BSC) and the Binary Erasure Channel (BEC).
Indeed, most of the techniques we have developed in classical information theory are tailored for memoryless channels, and fail spectacularly when applied to synchronization channels.

It is not hard to come up with examples of synchronization channels. Arguably, the simplest such model is the \emph{deletion channel}.
This channel receives a string of bits as input, and deletes each input bit independently with some deletion probability $d$.
The resulting output is a subsequence of the input string, and the loss of synchronization stems from the fact that the receiver, upon observing the $j$-th output bit, is uncertain about its true position $i$ in the input string.

\begin{figure}
	\centering
	\includegraphics[width=0.5\textwidth]{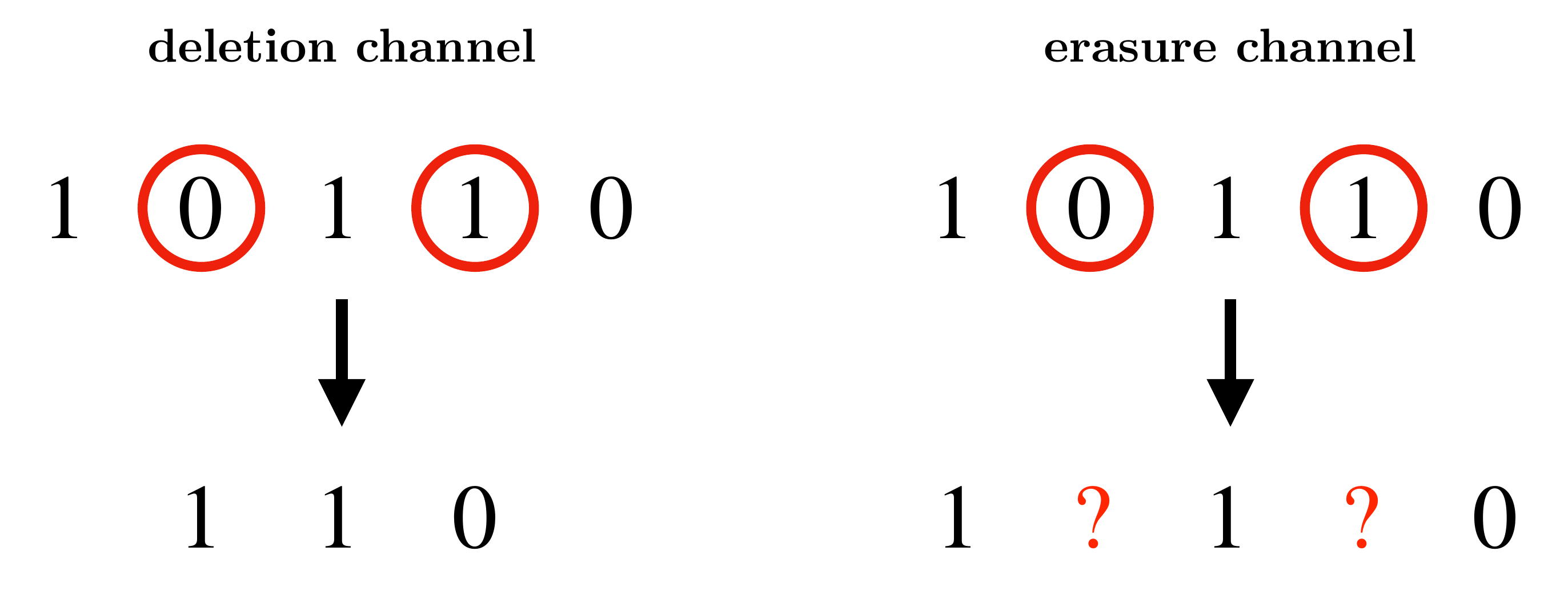}
	\caption{Comparison between the deletion and binary erasure channels. Circled bits are deleted or erased.}
	\label{fig:compdelbec}
\end{figure}

The deletion channel seems similar to the BEC, with the only difference being the fact that deleted bits are not replaced by a special symbol in the former (Figure~\ref{fig:compdelbec} illustrates the differences between the two models).
However, despite their similarities, it is much easier to analyze the BEC than the deletion channel.
In fact, although the capacity of the BEC with erasure probability has been known to equal $1-d$ for more than 70 years~\cite{Sha48} (achieved by a uniform input distribution), the capacity of the deletion channel is still unknown, although it is trivially upper bounded by $1-d$ by a
reduction that removes erased symbols incurred by a BEC.
As an example of the inadequacy of current information-theoretic techniques for understanding the deletion channel, we remark that a uniformly random codebook, which works very well for a large class of memoryless channels (including the BSC and BEC), performs badly under i.i.d.\ deletions (save for the low deletion probability regime, as we will see later).

Other well-studied synchronization errors include \emph{replications}, where an input bit is replaced by several copies of its value. Depending on the setting, the number of replications may be picked adversarially or independently according to some distribution over the integers (e.g., geometric or Poisson replications).
A class of synchronization errors that does not fit in the two families already discussed are \emph{insertions}.
In this case, a uniformly random symbol is added \emph{after} the input symbol in the string.
These types of errors and associated channels are discussed in a more rigorous way in Section~\ref{sec:othertypes}.

Besides being a source of fundamental open problems in information theory and theoretical computer science, channels with deletions, replications, and insertions appear naturally in several practical scenarios.
These include magnetic and optical data storage~\cite{KV93,Bou94}, multiple sequence alignment in computational biology~\cite{Lev01,Lev01b,BKKM04}, document exchange~\cite{Orl91,Hae18,CJLW18}, and, more recently, DNA-based data storage systems~\cite{CGK12,YKG+15,YGM17,OAC+18,CGMR19,BLS19}, racetrack memories~\cite{TSK+11,CKVVY18,CGVVY18,SB19}, and bit-patterned magnetic recording~\cite{KV11,AAA+15}.

Moreover, other than the direct applications above, the capacity of synchronization channels is also intimately connected to other problems in information theory and theoretical computer science.
Examples include the complexity of estimating the edit (also known as {Levenshtein}) distance between two strings in several settings~\cite{AK10,CDGKS18}, low-distortion embeddings of edit distance into other norms~\cite{OR07}, estimates on the expected length of the longest common subsequence between two random strings~\cite{KLM05}, the number of subsequences generated by deletion patterns~\cite{MKB08,LL15}, and cryptography~\cite{DORS08}.


The study of the fundamental limits of, and codes for, synchronization channels began with the seminal works of Gallager~\cite{Gal61}, Levenshtein~\cite{Lev65, Lev65b}, Dobrushin~\cite{Dob67}, Ullman~\cite{Ull67}, and Zigangirov~\cite{Zig69}.
This included both channels with i.i.d.\ and adversarial deletions, replications, and insertions.
The main focus of this survey lies on the fundamental limits of communication through channels with deletions, insertions, and replications with a constant rate of (both i.i.d.\ and adversarial) errors, and both in the regimes of vanishing and zero decoding error probability.
We present not only the main results in each specific topic, but also give a bird's eye (sometimes at different altitudes) of the ideas and techniques behind most results.
Some of the results we cover here can also be found in Mitzenmacher's survey~\cite{Mit09}.
After more than a decade past \cite{Mit09}, many new results and techniques have appeared.
We discuss these new contributions, and attempt to give a novel perspective on older results.

Although we do not discuss explicit constructions of codes robust against synchronization errors in detail, we remark that this is still a very active research area.
Indeed, even in the very basic case with a small number of adversarial deletions, first studied by Levenshtein~\cite{Lev65}, many important problems, with connections to combinatorics and number theory, remain open. Sloane's survey~\cite{Slo02} provides a great overview of this elegant setting.
Furthermore, a very complete account of coding schemes developed for various models with synchronization errors can be found in the survey by Mercier, Bhargava, and Tarokh~\cite{MBT10}.

\subsection{Some types of synchronization channels}\label{sec:othertypes}

In this section, we give a more careful overview of some types of synchronization channels that we will be focusing on in this survey.

\paragraph{Repeat channels}

\emph{Repeat channels} are a natural generalization of the deletion channel. These are channels that replicate each input symbol $x_i$ a total of $R_i$ times consecutively in the output (where $R_i=0$ means $x_i$ is deleted), where $R_i$ are i.i.d.\ according to some replication distribution $R$ over $\{0,1,2,\dots\}$.
Observe that the deletion channel can be seen as a replication channel with $R=\Ber_{1-d}$, where $\Ber_{1-d}$ denotes a Bernoulli distribution with success probability $1-d$.
In other words, $\Pr[R=0]=d$ and $\Pr[R=1]=1-d$.

Other notable repeat channels that introduce deletions which have been studied in the literature include the Poisson-repeat channel and the geometric deletion channel.
For the Poisson-repeat channel we have $R=\Poi_\lambda$, where $\Poi_\lambda$ denotes a Poisson distribution with mean $\lambda$. In this case, we have
\begin{equation*}
    \Pr[R=r]=\frac{e^{-\lambda}\lambda^r}{r!},\quad r=0,1,2,\dots,
\end{equation*}
and the deletion probability is $\Pr[R=0]=e^{-\lambda}$.
For the geometric deletion channel we have $R=\Geom_p$, where $\Geom_p$ denotes a geometric distribution with success probability $p$. In this case,
\begin{equation*}
    \Pr[R=r]=(1-p)^r p,\quad r=0,1,2,\dots,
\end{equation*}
and the deletion probability is $\Pr[R=0]=p$.

Throughout this work, we will denote the capacity of a repeat channel with replication distribution $R$ by $C(R)$.
In the special case of the deletion channel with deletion probability $d$, we denote its capacity by $C(d)$.
We remark that $C(R)$ is not known for any non-trivial replication distribution $R$.

\paragraph{Sticky channels and run-length encoding}

Repeat channels that do not introduce deletions (i.e., $\Pr[R=0]=0$) are called \emph{sticky channels}.
In a sense, sticky channels are the easiest type of repeat channels to analyze, and they are also connected to practical applications (e.g., see~\cite{FMS09,MDSG16}).
Widely studied examples of such channels include the duplication channel, which independently duplicates each input bit with probability $p$ (i.e., $R=1+\Ber_{p}$), and the geometric sticky channel, which independently replicates each input bit according to $R=1+\Geom_p$.
In both cases, we call $p$ the \emph{replication parameter}.

\begin{figure}
	\centering
	\includegraphics[width=0.5\textwidth]{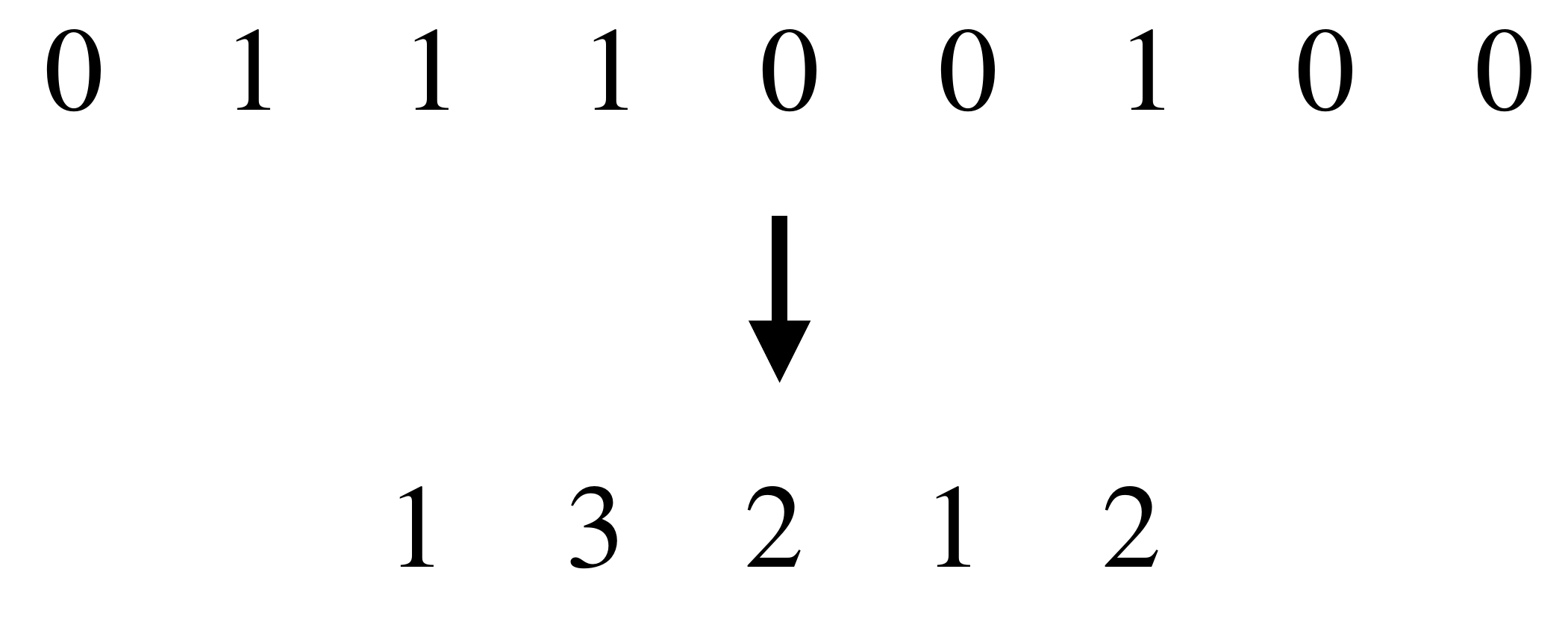}
	\caption{The run-length encoding of a binary string.}
	\label{fig:runlength}
\end{figure}

We proceed to explain why sticky channels are the easiest repeat channels.
First, in general it is useful to represent the input to a repeat channel by its \emph{run-length encoding}.
More precisely, suppose our input string for the repeat channel is
\begin{equation*}
    x=0^{\ell_1} 1^{\ell_2} 0^{\ell_3}\dots,
\end{equation*}
where the different $0^{\ell_i}$ and $1^{\ell_j}$ are called \emph{runs} of $x$. Then, the run-length encoding of $x$ is
\begin{equation*}
    (0,\ell_1,\ell_2,\ell_3,\dots).
\end{equation*}
For the particular application of studying the capacity of repeat channels, we may without loss of generality assume that every input string $x$ starts with a $0$.
This does not affect the capacity of the channel, and allows to use the simpler run-length encoding
\begin{equation*}
    (\ell_1,\ell_2,\ell_3,\dots)
\end{equation*}
for $x$, with the understanding that odd numbered runs correspond to $0$'s and even numbered runs correspond to $1$'s.
Figure~\ref{fig:runlength} depicts the run-length encoding of a particular string.

The behavior of repeat channels under run-length encoding is easy to describe. Observe that each input run of length $\ell$ is independently mapped to an output run of length
\begin{equation*}
    R^{(\ell)}=\sum_{i=1}^\ell R_i,
\end{equation*}
where the $R_i$ are i.i.d.\ according to $R$. Output runs of length $0$ are simply omitted from the output.
Under some choices of $R$, the sum $R^{(\ell)}$ has a nice structure.
For example, if $R=\Ber_{1-d}$, as is the case for the deletion channel, then $R^{(\ell)}=\Bin_{\ell,1-d}$, where $\Bin_{\ell,1-d}$ denotes a binomial distribution with $\ell$ trials and success probability $1-d$.
Moreover, if $R=\Geom_p$, then $R^{(\ell)}=\NB_{\ell,p}$, where
\begin{equation*}
    \NB_{\ell,p}(y)=\binom{y+\ell-1}{y}(1-p)^\ell p^y,\quad y=0,1,2,\dots
\end{equation*}
is the negative binomial distribution with $\ell$ failures and success probability $p$.

For the special case of the sticky channels, we have $\Pr[R^{(\ell)}=0]=0$ for every $\ell\geq 1$.
In other words, no input run is ever deleted.
This has significant implications. 
Notably, in order to determine the capacity of a sticky channel with replication distribution $R$, it suffices to understand the Discrete \emph{Memoryless} Channel (DMC) that maps integers $\ell\geq 1$ to $R^{(\ell)}$.
Observe that a sticky channel is memoryless between runs and we ``pay'' $\ell$ (out of total input length) to send a run of length $\ell$ through the DMC.
Therefore, the capacity of this sticky channel, $C(R)$, is equal to the \emph{capacity per unit cost} of the DMC $\Ch'$ that outputs $R^{(\ell)}$ on input $\ell\in\{1,2,\dots\}$.
In other words, we have
\begin{equation}\label{eq:stickycapunitcost}
    C(R) = \max_L\frac{I(L;R^{(L)})}{\E[L]},
\end{equation}
where the maximum is taken over all distributions $L$ supported on $\{1,2,\dots\}$.
The relationship in~\eqref{eq:stickycapunitcost} makes sticky channels easier to study.
Nevertheless, determining the exact capacity of any non-trivial sticky channel is still an interesting open problem.


\paragraph{Insertion channels}

An insertion error occurs when a \emph{random} bit is inserted into the string after a given input bit.
Observe that these are different errors than replications, where the inserted bits are copies of the input bit.
One may also think of insertions as complementary to deletions, in the sense that a deletion followed by an insertion makes a substitution error (i.e., a bit-flip).
Moreover, as we shall see, substitution errors are harder to decode from than deletions or insertions.

With a bit more care, we say there is a single insertion after input bit $x_i$ if $x_i$ is replaced in the input string by $x_i0$ with probability $1/2$ and $x_i 1$ with probability $1/2$.
An insertion channel with insertion probability $p_i$ independently corrupts each input bit with an insertion error with probability $p_i$.
We note that some works have considered channels that insert several random bits after an input bit (dictated by some distribution over the non-negative integers), and also channels
that combine deletions, replications, and insertions.

\subsection{Organization}

This survey is organized as follows: In Section~\ref{sec:shannontype}, we discuss the equality of the channel and information capacities for synchronization channels, along with related results.
Sections \ref{sec:lb}~and~\ref{sec:ub} are dedicated to capacity lower and upper bounds for synchronization channels, respectively.
The regimes of small and high deletion probabilities for the deletion channel are discussed in Section~\ref{sec:limit}.
The capacity of synchronization channels affected by memoryless errors is discussed in Section~\ref{sec:noisy}.
In Section~\ref{sec:multiuse}, we consider the multi-use setting for the deletion channel.
Finally, in Section~\ref{sec:zerorate} we study the zero-rate threshold for some adversarial synchronization channels.

\subsection{Notation}

Random variables are usually denoted by uppercase letters such as $X$, $Y$, and $Z$, and we may confuse a random variable with its distribution where appropriate. The support of a random variable $X$ is denoted by $\supp(X)$.
We write $X\rightarrow Y\rightarrow Z$ to say that $X$, $Y$, and $Z$ form a Markov chain (in this order).
We may also write $X\leftarrow \cS$ to mean that $X$ is sampled uniformly at random from the set $\cS$.
We denote the base-2 logarithm by $\log$ and the binary entropy function by $h(p)=-p\log p-(1-p)\log(1-p)$.
The natural logarithm is denoted by $\ln$.
The (Shannon) entropy of a random variable $X$ is denoted by $H(X)$, and $I(X;Y)$ denotes the mutual information between $X$ and $Y$. 
The Kullback-Leibler (KL) divergence between $X$ and $Y$ is denoted by $\KL(X||Y)$.
Unless otherwise stated, capacities are presented in bits/channel use.
Given a string $x\in\Sigma^n$, we say a string $y\in\Sigma^m$ is a \emph{subsequence} of $x$ if there exist indices $1\leq i_1<i_2<\cdots<i_m\leq n$ such that $x_{i_j}=y_j$. Moreover, we say $y$ is a \emph{substring} of $x$ if $i_j=i_1+j-1$ for all $j=1,\dots,m$. A \emph{run of length $\ell$} in a string $x\in\Sigma^n$ is a substring $y=s^\ell$ for some $s\in\Sigma$ that cannot be extended.
Given a vector or string $x$, we denote its length by $|x|$.
For two strings $x$ and $y$, we say $z$ is a common subsequence of $x$ and $y$ if it appears as a subsequence in both $x$ and $y$.
We denote by $\LCS(x,y)$ the \emph{length} of the longest common subsequence of $x$ and $y$.
We denote $\Sigma^*=\bigcup_{n=0}^\infty \Sigma^n$, where $\Sigma^0$ contains only the empty string.

\paragraph{Discrete channels}

In this survey, we will be dealing solely with discrete channels.
Such a channel $\Ch$ maps elements of $\cX^*$ to elements of $\cY^*$, and is characterized by a conditional probability distribution $p(\cdot|x)$ for every $x\in\cX^*$.
We call $\cX$ and $\cY$ the input and output alphabets of $\Ch$, respectively.
We denote the output distribution of $\Ch$ given input $x\in\cX^*$ by $Y_x$. Then, we have $\Pr[Y_x=y]=p(y|x)$.
For an input distribution $X$, we denote its corresponding output distribution under $\Ch$ by $Y_X$.
In other words, $Y_X$ satisfies $\Pr[Y_X=y]=\sum_{x\in\cX} \Pr[X=x]\cdot \Pr[Y_x=y]$.

We say such a channel $\Ch$ with associated conditional probability distribution $p$ is a \emph{Discrete Memoryless Channel} (DMC) if it maps $\cX^n$ to $\cY^n$ and
\begin{equation*}
    p(y_1,\dots,y_n|x_1,\dots,x_n)=\prod_{i=1}^n p(y_i|x_i).
\end{equation*}
As a result, in order to analyze a DMC $\Ch$, it suffices to study its behavior on inputs $x\in\cX$.
Well-known examples of DMC's include the Binary Symmetric Channel (BSC) and the Binary Erasure Channel (BEC).
We stress that synchronization channels are not DMC's.




\section{A Shannon-type theorem for the channel capacity}\label{sec:shannontype}

The focus of this survey lies on the capacity of synchronization channels. Here, we mostly mean capacity in the usual sense: The supremum of all rates at which \emph{reliable} transmission is possible (i.e., with vanishing error probability as the block length increases). More precisely, we have the following definition.

\begin{defn}[Achievable rate and capacity of a channel]
    Given a channel $\Ch$ with input alphabet $\cX$ and output alphabet $\cY$, we say a real number $R\geq 0$ is an \emph{achievable rate for $\Ch$} if for every $n$ large enough there exists a codebook $\cC\subseteq \cX^n$ of size $|\cC|=\lceil|\cX|^{Rn}\rceil$ and a function $\Dec:\cY^*\to\cC$ such that
    \begin{equation*}
        \Pr[\Dec(Y_x)=x]\geq 1-\lambda(n)
    \end{equation*}
    for some $\lambda(n)\to 0$ and all $x\in\cC$, where $Y_x$ denotes the output distribution of $\Ch$ on input $x$.
    
    Then, the \emph{capacity of $\Ch$}, denoted by $\Ca(\Ch)$, is given by
    \begin{equation*}
        \Ca(\Ch)=\sup\{R\geq 0: \textrm{ $R$ is an achievable rate for $\Ch$}\}.
    \end{equation*}
\end{defn}

In his seminal work, Shannon~\cite{Sha48} proved the \emph{noisy channel coding theorem}: If $\Ch$ is a discrete memoryless channel (DMC), then its capacity allows the following characterization.
\begin{thm}[\cite{Sha48}]\label{thm:cap}
    Suppose $\Ch$ is a DMC. Then, the capacity of $\Ch$ satisfies
    \begin{equation}\label{eq:charcap}
        \Ca(\Ch)=\lim_{n\to\infty}\max_{X^{(n)}}\frac{I(X^{(n)};Y_{X^{(n)}})}{n}=\max_X I(X;Y_X).
    \end{equation}
    In the middle expression of~\eqref{eq:charcap}, the maximum inside the limit is taken over all distributions $X^{(n)}$ over $\cX^n$, and $Y_{X^{(n)}}$ is the corresponding output distribution.
    The maximum in the right-hand side expression is taken over all distributions $X$ over $\cX$, and $Y_X$ is the associated output distribution of $\Ch$ with input $X$.
\end{thm}
In fact, it is possible to prove a \emph{strong converse} to Shannon's noisy channel coding theorem~\cite{Wol57}: If we attempt to communicate at rates exceeding capacity, then not only is the error probability bounded away from $0$, but it actually converges to $1$ as the block length $n$ increases.

Synchronization channels, however, are not memoryless. Therefore, it is not clear whether an analogue of~\eqref{eq:charcap} holds for them. 
This was settled by Dobrushin~\cite{Dob67}, who showed that this is the case for a large class of synchronization channels which includes repeat channels with well-behaved replication distributions.
Consider a synchronization channel $\Ch$ with input and output alphabets $\cX=[a]$ and $\cY=[b]^*$. When $x\in\cX^*$ is sent through $\Ch$, each input symbol $x_i\in\cX$ is independently mapped to $y_i\in\cY$ according to a conditional probability distribution $p(\cdot|x_i)$, and the $y_i$'s are then concatenated (note that some $y_i$'s may be the empty string, which represents a deletion). The following holds.
\begin{thm}[\cite{Dob67}]\label{thm:capsynch}
    Let $\Ch$ be a synchronization channel such that for real constants $0<c_1\leq c_2$ it holds that
    \begin{equation}\label{eq:constdob}
        c_1\leq \sum_{y\in\cY} |y|\cdot p(y|x)\leq c_2
    \end{equation}
    for all $x\in\cX$. Then, we have
    \begin{equation}\label{eq:charcapsynch}
        \Ca(\Ch)=\lim_{n\to\infty}\max_{X^{(n)}}\frac{I(X^{(n)};Y_{X^{(n)}})}{n},
    \end{equation}
    where the maximum is taken over all $X^{(n)}$ supported on $\cX^n$, and $Y_{X^{(n)}}$ is the associated output distribution.
    Moreover, the capacity is achieved by a stationary ergodic input source.
\end{thm}

The characterization in~\eqref{eq:charcapsynch} does not seem to be very helpful in determining the exact capacity of synchronization channels.
Nevertheless, it has been useful in the derivation of good capacity lower bounds for such channels.
Moreover, the fact that we can restrict ourselves to stationary ergodic input sources has been crucial in the derivation of capacity upper bounds.

Analogously to the memoryless case, a strong converse to Theorem~\ref{thm:capsynch} was independently proved by Ahlswede and Wolfowitz~\cite{AW71} and Kozglov~\cite{Koz71} under a different assumption than~\eqref{eq:constdob}. 
Namely, the strong converse requires that there exists a constant $M$ such that
\begin{equation}\label{eq:constsynch2}
    \textnormal{$p(y|x)>0$ holds for some $x\in\cX$ only if $|y|\leq M$.}
\end{equation}
In particular, repeat channels with replication distributions having unbounded support, such as the geometric sticky channel and the Poisson-repeat channel, do not satisfy this constraint. 
Ahlswede and Wolfowitz~\cite{AW71} go farther and also prove strong converses for other models, such as synchronization channels with feedback.
Although not our main focus, we note that Theorem~\ref{thm:capsynch} has been generalized to continuous channels by Stambler~\cite{Sta70}, and to channels with timing errors and inter-symbol interference by Zeng, Mitran, and Kav\v{c}i\'{c}~\cite{ZMK06}.

When attempting to bound or approximate the capacity of a synchronization channel $\Ch$, it is useful to consider the relationship between $\Ca(\Ch)$ and the capacity at ``finite block length" $\Ca_n(\Ch)$, defined as
\begin{equation}\label{eq:finitelength}
    \Ca_n(\Ch)=\frac{1}{n}\max_{X^{(n)}}I(X^{(n)};Y_{X^{(n)}}),
\end{equation}
where the maximum is taken over all distributions $X^{(n)}$ supported on $\cX^n$.
Theorem~\ref{thm:capsynch} states that
\begin{equation*}
    \Ca(\Ch)=\lim_{n\to\infty} \Ca_n(\Ch).
\end{equation*}
However, one can prove a stronger statement via Fekete's lemma for \emph{subadditive} sequences $(a_n)_{n=1}^\infty$ of real numbers, which satisfy $a_{m+n}\leq a_m+a_n$ for all $m,n\geq 1$.
\begin{lem}[Fekete's lemma~\cite{Fek23}]\label{lem:fekete}
    Suppose $(a_n)_{n=1}^\infty$ is a subadditive sequence of real numbers. Then, we have
    \begin{equation*}
        \lim_{n\to\infty} \frac{a_n}{n} = \inf_n \frac{a_n}{n}.
    \end{equation*}
\end{lem}

Fekete's lemma can be used to prove the following result.
\begin{thm}\label{thm:finitecap}
    We have
    \begin{equation*}
        \Ca(\Ch)=\inf_n \Ca_n(\Ch).
    \end{equation*}
    In particular, it holds that $\Ca_n(\Ch)\geq \Ca(\Ch)$ for all $n\geq 1$.
\end{thm}

    By Fekete's lemma, it suffices to prove that the sequence $(n\cdot \Ca_n(\Ch))_{n=1}^\infty$ is subadditive. For arbitrary $m,n\geq 1$, consider the modified channel $\Ch'$ obtained by adding a marker between the first $n$ outputs of $\Ch$ and the remaining $m$ outputs. It holds that $\Ch$ is a degraded version of $\Ch'$ (obtained by removing the marker). 
    
    Consider an arbitrary input source $X$ over $\cX^{n+m}$. Let $X_1$ denote its restriction to the first $n$ symbols, and $X_2$ its restriction to the last $m$ symbols. If $Y_i$ denotes the output of $\Ch$ under input $X_i$ for $i=1,2$ and $Y$, $Y'$ denote the output of $\Ch$ and $\Ch'$ under input $X$, respectively, we have
    \begin{align*}
        I(X;Y)&\leq I(X;Y')\\
        &\leq I(X_1;Y_1)+I(X_2;Y_2).
    \end{align*}
    Subadditivity follows easily from this inequality by noting that $X$ is arbitrary.

The rate at which $\Ca_n(\Ch)$ converges to $\Ca(\Ch)$ has also been characterized for synchronization channels. Taking into account Theorem~\ref{thm:finitecap}, it is known~\cite{Dob67} that
\begin{equation}\label{eq:sandwichcap}
    \Ca_n(\Ch)-\frac{\log(n+1)}{n}\leq \Ca(\Ch)\leq \Ca_n(\Ch)
\end{equation}
for all $n\geq 1$. Moreover, it has also been shown that the left-hand side inequality in~\eqref{eq:sandwichcap} is tight up to a multiplicative constant~\cite{AW71}.
As we discuss later on in Section~\ref{sec:ub}, one could possibly use~\eqref{eq:sandwichcap} coupled with numerical algorithms to potentially derive good capacity bounds for synchronization channels.
However, this turns out to be computationally infeasible for decent values of $n$.

\paragraph{Capacity per unit cost}
As already mentioned in Section~\ref{sec:othertypes}, we will also need to work with the capacity \emph{per unit cost} of DMC's with real-valued input later on.
We proceed to define it.
\begin{defn}[Capacity per unit cost]\label{def:capcost}
    Given a DMC $\Ch$ with input and output alphabets $\cX\subseteq \mathbb{R}$ and $\cY$, respectively, its \emph{capacity per unit cost with cost function $c$}, denoted by $\overline{\Ca}_c(\Ch)$, is given by
    \begin{equation*}
        \overline{\Ca}_c(\Ch)=\max_X \frac{I(X;Y_X)}{\E[c(X)]},
    \end{equation*}
    where the maximum is over all possible input distributions $X$ over $\cX$, and $Y_X$ denotes the associated output distribution.
    
    If $c(x)=x$ for all $x$, we simply write  $\overline{\Ca}(\Ch)$ for the capacity per unit cost of $\Ch$.
\end{defn}




\section{General capacity lower bounds for synchronization channels}\label{sec:lb}

In this section, we give an account of the development of capacity lower bounds for synchronization channels (mainly for repeat channels) and the underlying techniques. Some of these bounds have already been discussed by Mitzenmacher~\cite{Mit09}. However, the topic has developed since then, and we discuss some more recent work.
For the sake of clarity, we will center our exposition mainly around the (often simpler) deletion channel.
We remark, however, that this will not always be the case, as some techniques are tailored for other types of repeat channels (in particular, the tight lower bounds for sticky channels).

The first capacity lower bound for the deletion channel was derived by Gallager~\cite{Gal61} and Zigangirov~\cite{Zig69} (who also considered a channel combining deletions with geometric insertions) by considering the performance of convolutional codes under synchronization errors. They showed that
\begin{equation}\label{eq:lbBSC}
    C(d)\geq 1-h(d),
\end{equation}
where $C(d)$ denotes the capacity of the deletion channel with deletion probability $d$. Ullman~\cite{Ull67} studied the \emph{zero-error} capacity of the deletion channel, and derived a capacity lower bound in that setting. However, given that the zero-error setting is much more demanding than the vanishing error setting we consider in most of this survey, his lower bound is generally more pessimistic than~\eqref{eq:lbBSC}.

There is an alternative and simpler proof of~\eqref{eq:lbBSC} that fits our current perspective on capacity lower bounds for synchronization channels better. When attempting to lower bound the capacity of a repeat channel, it is natural to consider the rate achieved by a uniformly random codebook (i.e., a uniform channel input distribution) with an appropriate decoder. Using this approach, Diggavi and Grossglauser~\cite{DG06} re-derived~\eqref{eq:lbBSC}.
The decoder considered is simple: Given the output $y$ of the deletion channel, the receiver verifies whether $y$ is a subsequence of only one codeword, and outputs that codeword. If this is not the case, then the receiver simply declares an error.

While~\eqref{eq:lbBSC}, which we saw is achieved by a random codebook, turns out to behave well for small $d$ (this is discussed in more detail in Section~\ref{sec:lowdel}), it degrades quickly as $d$ increases, and is trivial for $d\geq 1/2$.
This is not surprising, considering that deletions are not memoryless errors.
Therefore, one expects that a good input distribution for the deletion channel should also have memory.
One reasonable way to take this into account is to consider Markov chains as input distributions.
This was done as early as $1968$, when Vvedenskaya and Dobrushin~\cite{VD68} estimated the rates achieved over the deletion channel by Markov chains of order at most $2$ via numerical simulations, although their results cannot be assumed to be reliable~\cite{DM06}. Exploiting Monte-Carlo methods, these rates were later estimated again via numerical simulations by Kav\v{c}i\'c and Motwani~\cite{KM04} (also for channels with insertions).
We remark that both these works do not yield rigorous bounds, as their reported results are simulation-based.
Nevertheless, they provide a good picture of the true achievable rates, and they strongly suggest Markov input sources behave significantly better than a uniform input.

Using an arbitrary Markov chain of order $1$ as input and the same simple decoding procedure detailed above, Diggavi and Grossglauser~\cite{DG06} derived an analytical lower bound that improves on~\eqref{eq:lbBSC}. More precisely, they showed that
\begin{equation}\label{eq:markov1lb}
    C(d)\geq \frac{1}{\ln 2}\cdot \sup_{\gamma>0,0<p<1}[-(1-d)\ln((1-q)A+qB)-\gamma],
\end{equation}
where $q=1-\frac{1-p}{1+d(1-2p)}$, $A=\frac{(1-p)e^{-\gamma}}{1-p e^{-\gamma}}$, and $B=\frac{(1-p)^2e^{-2\gamma}}{1-p e^{-\gamma}}+pe^{-\gamma}$. Intuitively, we obtain~\eqref{eq:markov1lb} by optimizing the achievable rate over order $1$ Markov chains over $\bits$ with transition probability $1-p$ from $0$ to $1$ and vice-versa.
A uniform input distribution corresponds to $p=1/2$.
Observe that we may think of a Markov chain of order $1$ as an input distribution with runs following a geometric distribution (starting at $1$).

The lower bound in~\eqref{eq:markov1lb} was improved by Drinea and Mitzenmacher~\cite{DM06, DM07}, and the approach was generalized to other repeat channels. They also consider Markov chains of order $1$ as input distributions (equivalently, input distributions with i.i.d.\ geometric runs), but use a more careful decoding procedure, which they term \emph{jigsaw decoding}.
The main ideas behind jigsaw decoding are described extremely well in Mitzenmacher's survey~\cite{Mit09}.

Notably, Mitzenmacher and Drinea~\cite{MD06} exploit jigsaw decoding-based capacity lower bounds for the Poisson-repeat channel, where each input bit is replicated according to a $\Poi_\lambda$ distribution, and a connection between this channel and the deletion channel to derive the simple-looking lower bound
\begin{equation*}
    C(d)\geq 0.1185(1-d)>\frac{1-d}{9}
\end{equation*}
for all $d$.
This bound is especially relevant in settings where $d$ is close to $1$. We discuss it in more detail in Section~\ref{sec:highdel}.

A radically different approach towards capacity lower bounds was proposed by Kirsch and Drinea~\cite{KD10}, although with some ties to ideas used in~\cite{DM06,DM07}. Instead of considering a codebook generated by some distribution and the rate achieved under a specific decoding algorithm, they undertake a purely information-theoretic approach using Dobrushin's characterization of the capacity of repeat channels ( recall Theorem~\ref{thm:capsynch}).
In other words, for given input distributions $X_n$ over $n$-bit inputs, $n=1,2,\dots$, Kirsch and Drinea directly lower bound the \emph{information rate}
\begin{equation*}
    \lim_{n\to\infty} \frac{I(X^{(n)};Y_{X^{(n)}})}{n},
\end{equation*}
where $X^{(n)}$ is supported on $\bits^n$.
Any such lower bound directly yields a capacity lower bound for the repeat channel under consideration.
In particular, they consider $n$-bit input distributions $X^{(n)}$ generated by i.i.d.\ runs with arbitrary run-length distribution $P$,\footnote{To sample the input string $X^{(n)}$,
runs are generated according to $P$ until there are at least $n$ bits in the string. Then, the last run is truncated so that $|X^{(n)}|=n$. Observe that this introduces dependencies between different input run-lengths. However, we may assume that all runs in $X^{(n)}$ are i.i.d., as this will have no effect on the rate when $n\to\infty$ (assuming $P$ is well-behaved).} and show that
\begin{equation*}
    \lim_{n\to\infty} \frac{I(X^{(n)};Y_{X^{(n)}})}{n} = [\textrm{rate achieved by jigsaw decoding}]+\eps_P,
\end{equation*}
where $\eps_P$ is a positive term that depends on the input run-length distribution $P$. This term has the nice added property that
\begin{equation*}
    \eps_P=\lim_{m\to\infty}\eps_{P,m}
\end{equation*}
for a positive, non-decreasing, bounded sequence $(\eps_{P,m})_{m=1,2,\dots}$.
Therefore, it suffices to approximate (or even lower bound) $\eps_{P,m}$ for some $m$ appropriately to derive improved capacity lower bounds.
However, naive computation of $\eps_{P,m}$ requires dealing with many nested summations~\cite{KD10}, which makes this task infeasible in practice.

Summing up, on the positive side the strategy from~\cite{KD10} yields computable lower bounds that potentially beat (or match) \emph{any} lower bound obtained by considering codebooks generated by distributions with i.i.d.\ runs. On the other hand, as discussed above, it is currently practically infeasible to numerically approximate or lower bound $\eps_P$ appropriately. Therefore, Kirsch and Drinea resort to simulation-based techniques to estimate it for some range of parameters.
Their results strongly suggest (although without rigorous proof) that considering the extra term $\eps_P$ in the lower bound leads to significantly improved lower bounds when compared to jigsaw decoding.
The main ideas behind this result have been described extremely well in~\cite{Mit09}, and we refrain from doing so here.

Later, improved explicit lower bounds on the capacity of channels with deletions and insertions were derived by Venkataramanan, Tatikonda, and Ramchandran~\cite{VTR13}.
First, they only consider order-$1$ Markov chains as input distributions as in~\cite{DM06,DM07} (instead of any distribution with i.i.d.\ runs as in~\cite{KD10}).
However, similarly to the work of Kirsch and Drinea~\cite{KD10}, they work directly with the mutual information, and split the information capacity as
\begin{equation}\label{eq:splitsubopt}
    \lim_{n\to\infty}\frac{I(X^{(n)};Y_{X^{(n)}})}{n}=[\textrm{rate of sub-optimal decoder}]+\eps_P,
\end{equation}
where $\eps_P$ is a positive term that depends on the runlength distribution $P$ (which, in this case, follows a geometric distribution) used to generate the $X^{(n)}$.
Intuitively, if $X$ is the input distribution and $Y$ is the corresponding output distribution, this sub-optimal decoder (which is not explicit) is induced by enforcing that a certain random process $S$ (correlated with $X$ and $Y$) must be output as side information by the decoder. 
In other words, on input $Y$ the decoder must recover \emph{both} $X$ and $S$ with high probability, and the first term on the right-hand side of~\eqref{eq:splitsubopt} is the rate achieved by the \emph{best} decoder under this additional constraint.
We remark that a globablly optimal decoder does not necessarily have to be able to recover the side information $S$ besides $X$.


The strategy above allows the authors to lower bound each term on the right-hand side of~\eqref{eq:splitsubopt} (and hence the rate achieved by order-1 Markov chains) by an expression that can be numerically computed in practice. As a result, they improve upon the lower bounds from~\cite{DM06,DM07} for deletion probability $d\leq 0.3$.
We note that they also apply the same high-level strategy to derive capacity lower bounds for the insertion and ins/del channels.

We proceed to sketch the main ideas behind the result of~\cite{VTR13} for the deletion channel.
For a fixed $n$, let $X=X^{(n)}$ be the $n$-bit input distribution to the deletion channel and $Y=Y_X$ its corresponding output distribution with $M$ bits.
Then, consider the tuple of integers $S=(S_1,S_2,\dots,S_{M+1})$ where $S_i$ denotes the number of runs in $X$ that are deleted between the input bits corresponding to $Y_{i-1}$ and $Y_i$ ($S_1$ denotes the number of runs deleted before $Y_1$, and $S_{M+1}$ denotes the number of runs deleted after $Y_M$).
Then, using basic information-theoretic equalities, we have
\begin{equation*}
    H(X|Y) = H(X,S|Y)-H(S|X,Y).
\end{equation*}
Renaming the tuple $(X,Y,S)$ in the discussion above as $(X^{(n)},Y^{(n)},S^{(n)})$, we conclude that
\begin{align}
    \lim_{n\to\infty}\frac{I(X^{(n)};Y^{(n)})}{n}&=h(p)-\lim_{n\to\infty}\frac{H(X^{(n)}|Y^{(n)})}{n}\nonumber\\
    &=h(p)-\lim_{n\to\infty}\frac{H(X^{(n)},S^{(n)}|Y^{(n)})}{n}+\lim_{n\to\infty}\frac{H(S^{(n)}|X^{(n)},Y^{(n)})}{n},\label{eq:rewritemutinf}
\end{align}
where $p$ is such that runs of $X^{(n)}$ are generated i.i.d.\ according to $1+\mathsf{Geom}(p)$ (thus $h(p)$ is the entropy rate of the Markov process).
Remarkably, the middle term in~\eqref{eq:rewritemutinf} has an analytical expression. Therefore, to obtain a good computable capacity lower bound, it suffices to find an appropriate lower bound for the right-hand side term of~\eqref{eq:rewritemutinf}.
This turns out to be doable with significant effort by carefully restricting the averaging in the conditional entropy term $H(S^{(n)}|X^{(n)},Y^{(n)})$ with respect to $(X^{(n)},Y^{(n)})$ to certain terms $(X^{(n)}=x,Y^{(n)}=y)$ for which the distribution $(S^{(n)}|X^{(n)}=x, Y^{(n)}=y)$ can be more easily understood.

Fertonani and Duman~\cite{FD10} presented a simple approach that yields numerical capacity lower bounds for the deletion channel. These bounds are close to, but do not improve upon, the bounds from~\cite{DM07,VTR13}.
Nevertheless, we discuss it here both due to its simplicity, and also because their strategy is the basis for the work of Castiglione and Kav\v{c}i\'c~\cite{CK15} where the current best \emph{simulation-based} capacity lower bounds for the deletion channel are derived (they analyze a more general class of channels too). These simulation-based results are obtained by carefully estimating the rate achieved by Markov chains of order $3$ under the deletion channel. We stress that these are not true lower bounds, in the sense that there is no rigorous proof.

The strategy of Fertonani and Duman~\cite{FD10} will also be discussed from a slightly different perspective in Section~\ref{sec:ub}. In fact, its main goal was the derivation of good numerical capacity upper bounds for the deletion channel.

\begin{figure}
	\centering
	\includegraphics[width=0.4\textwidth]{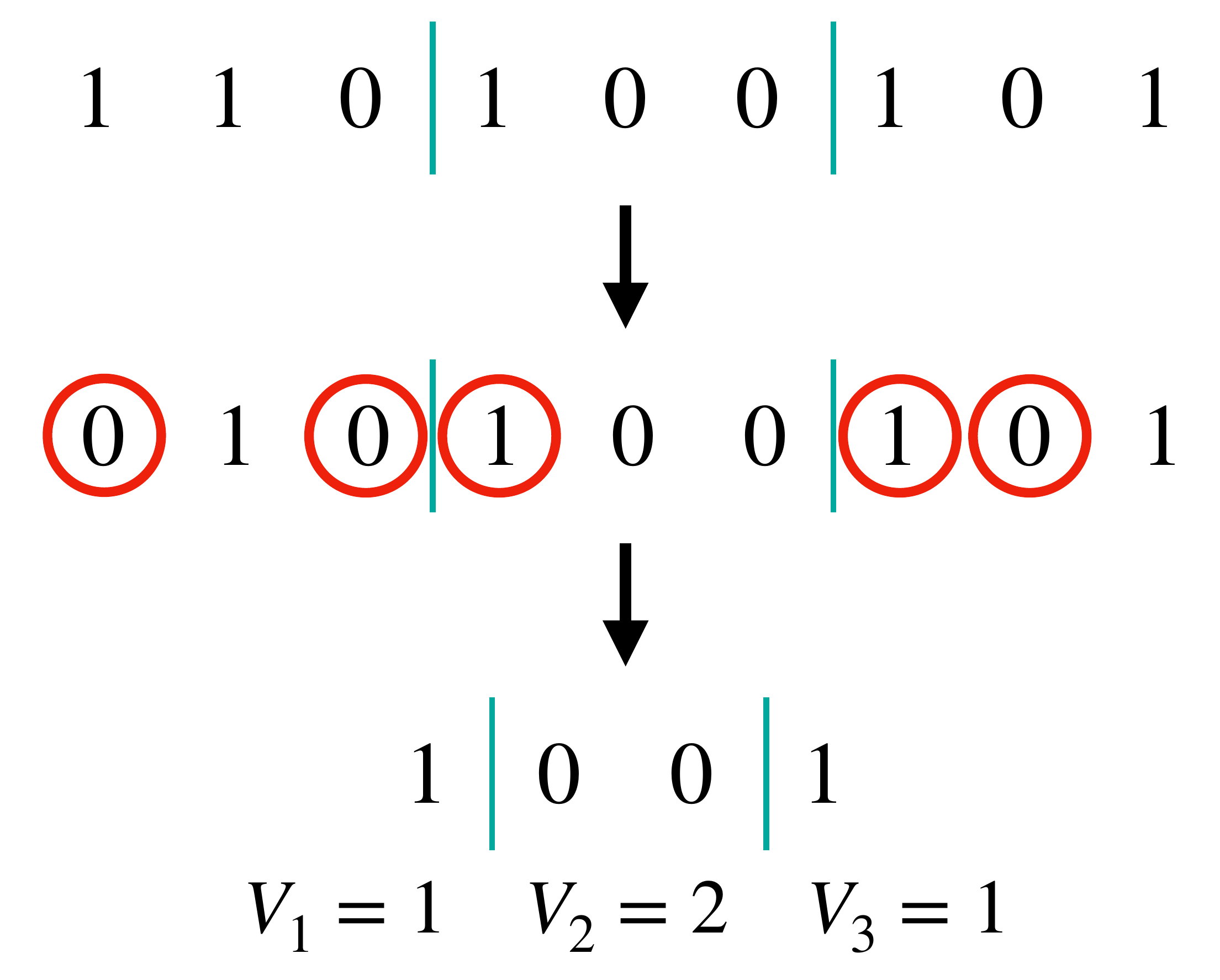}
	\caption{Adding undeletable markers to the input at regular intervals, and the resulting random process $V$. Circled bits are deleted.}
	\label{fig:regularmarkers}
\end{figure}

The bound in~\cite{FD10} is derived by adding undeletable markers to the input after every $\ell$ bits, for some constant $\ell$. More precisely, we can
consider the following random process $V$: Let $X$ be an $n$-bit input distribution for the deletion channel and $Y$ its associated output distribution. For some constant $\ell$ of our choice, partition $X$ into $n/\ell$ consecutive blocks $X^{[1]},X^{[2]},\dots,X^{[n/\ell]}$ of $\ell$ bits each, and let $Y^{[i]}$ denote the part of $Y$ coming from $X^{[i]}$. 
Then, we can define $V_i=|Y^{[i]}|$ and set $V=(V_1,\dots,V_{n/\ell})$.
This process is illustrated in Figure~\ref{fig:regularmarkers}.
Revealing $V$ as side information to the receiver and using basic information-theoretic inequalities immediately leads to the bounds
\begin{equation}\label{eq:basicmutinf}
    I(X;Y,V)\geq I(X;Y)\geq I(X;Y,V)-H(V).
\end{equation}
Since the $V_i$ are i.i.d., we conclude that
\begin{equation}\label{eq:basicent}
    \frac{1}{n}H(V)=\frac{1}{\ell}H(V_1).
\end{equation}
Note that the right-hand side of~\eqref{eq:basicent} is very easy to compute.
In fact, $V_1$ follows a $\Bin_{\ell,1-d}$ distribution.
Moreover, from the pair $(Y,V)$ we can fully determine all $Y^{[i]}$. As a result, the channel $X\mapsto (Y,V)$ is equivalent to $n/\ell$ independent copies of the channel $X^{[1]}\mapsto Y^{[1]}$. 
Therefore, we have
\begin{equation}\label{eq:genieV}
    \max_{X}\frac{1}{n}I(X;Y,V)=\max_{X^{[1]}}\frac{1}{\ell}I(X^{[1]};Y^{[1]})
\end{equation}
for every $n$.
Denoting the capacity of the channel that maps $X^{[1]}$ to $Y^{[1]}$ with \emph{fixed} input length $\ell$ (i.e., the right-hand side of~\eqref{eq:genieV}) by $C_\ell$, it follows 
from~\eqref{eq:genieV},~\eqref{eq:basicent}, and~\eqref{eq:basicmutinf}
that
\begin{align}\label{eq:capineqV}
    C_\ell \geq \lim_{n\to\infty}\max_{X^{(n)}}\frac{I(X^{(n)};Y_{X^{(n)}})}{n}\geq C_\ell-\frac{1}{\ell}H(V_1)
\end{align}
for all constants $\ell$.
By Theorem~\ref{thm:capsynch}, we conclude from~\eqref{eq:capineqV} that
\begin{align}\label{eq:cellbounds}
    C_\ell \geq C(d)\geq C_\ell-\frac{1}{\ell}H(V_1),
\end{align}
recalling $C(d)$ denotes the capacity of the deletion channel with deletion probability $d$.
If $\ell$ is a small enough constant, then $C_\ell$ can be approximated numerically to great accuracy in a suitable amount of time with recourse to the Blahut-Arimoto algorithm~\cite{Ari72,Bla72} (in~\cite{FD10}, the authors go up to $\ell=17$).

To finalize, we remark that, instead of following the reasoning above, one could attempt to directly exploit the well-known general relationship between $C_n(d)=\frac{1}{n}\max_{X^{(n)}} I(X^{(n)};Y_{X^{(n)}})$ (where $X^{(n)}$ is an $n$-bit input distribution to the deletion channel and $Y_{X^{(n)}}$ is the corresponding output) and $C(d)$ to obtain capacity bounds. Recalling~\eqref{eq:sandwichcap}, we have
\begin{equation}\label{eq:sandwichcapdel}
    C_n(d)\geq C(d)\geq C_n(d)-\frac{\log(n+1)}{n}
\end{equation}
for every $n$.
Observe that $C_n(d)$ can potentially be computed with the help of the Blahut-Arimoto algorithm.
However, this approach is computationally infeasible whenever $n$ is not very small.

\paragraph{Capacity lower bounds for sticky channels}

Several works have focused on obtaining capacity lower bounds for \emph{sticky} channels.
These channels appear to have a simpler structure than channels with deletions.
In fact, studying the capacity of a sticky channel boils down to analyzing the capacity \emph{per unit cost} of a certain DMC over the positive integers.
This opens the door to lower bound techniques that are not available or are harder to realize for channels with deletions and insertions.
Consequently, previous efforts in this topic have resulted in tight numerical lower bounds for many sticky channels.

As previously discussed in Section~\ref{sec:othertypes}, sticky channels, first studied on their own by Mitzenmacher~\cite{Mit08}, replicate each input bit independently according to a replication distribution $R$ over the \emph{positive} integers.
This means that no input bit is deleted.
In particular, it easily follows that the capacity of a sticky channel $\Ch$, denoted $\Ca(\Ch)$, equals the capacity per unit cost (with costs $c(x)=x$) of a DMC $\Ch'$ over the positive integers.
In other words, recalling Definition~\ref{def:capcost} we have
\begin{equation}\label{eq:equivcapunitcost}
    \Ca(\Ch)=\max_X \frac{I(X;Y_X)}{\E[X]}=\overline{\Ca}(\Ch'),
\end{equation}
where we maximize over all input distributions $X$ supported on $\{1,2,\dots\}$, and $Y_X$ denotes the output distribution of $\Ch'$ with input $X$.
Two notable examples of sticky channels are the duplication channel, where $R=1+\Ber_p$, and the geometric sticky channel, where $R=1+\Geom_p$.
For both these channels, the equivalent DMC over the positive integers has a nice form.
Indeed, the ``equivalent'' DMC $\Ch'$ for the duplication channel maps integers $x>0$ to $x+\Bin_{x,p}$, and the channel $\Ch'$ for the geometric sticky channel maps integers $x>0$ to $x+\NB_{x,p}$.

Mitzenmacher~\cite{Mit08} exploited the equivalence in~\eqref{eq:equivcapunitcost} to derive numerical capacity lower bounds for both the duplication and geometric sticky channels.
The capacity per unit cost of DMC's with \emph{finite} input and output alphabets and positive symbol costs can be numerically computed using a variant of the Blahut-Arimoto algorithm due to Jimbo and Kunisawa~\cite{JK79}. As a by-product, the Jimbo-Kunisawa algorithm also outputs the capacity-achieving distribution.
However, here we have to deal with a DMC $\Ch'$ with infinite input and output alphabets.
Nevertheless, this is easily dealt with if we are aiming for good lower bounds only.
Indeed, it suffices to consider a modified DMC $\Ch'_{a,b}$ obtained by truncating the input and output alphabets of $\Ch'$. More precisely, $\Ch'_{a,b}$ behaves exactly like $\Ch'$, but only accepts inputs $x\leq a$ and, if the output $y$ of $\Ch'$ satisfies $y>b$, then $\Ch'_{a,b}$ outputs a special symbol $\bot$ instead of $y$.\footnote{Observe that we do not need the second constraint (truncation of output alphabet) when dealing with the duplication channel, since in that case any truncation of the input alphabet induces a finite output alphabet. However, the same is not true for the geometric sticky channel.}
It follows easily from the definition of $\Ch'_{a,b}$ that
\begin{equation*}
    \overline{\Ca}(\Ch'_{a,b})\leq \overline{\Ca}(\Ch').
\end{equation*}
We can then apply the Jimbo-Kunisawa algorithm to compute $\overline{\Ca}(\Ch'_{a,b})$ numerically for sufficiently small $a$ and $b$.

As determined by Mitzenmacher~\cite{Mit08} and later Mercier, Tarokh, and Labeau~\cite{MTL12}, it turns out the simple strategy above already yields tight numerical capacity lower bounds for both the duplication and geometric sticky channels over the full range of the replication parameter $p$.
Moreover, as evidenced in~\cite{MTL12}, codebooks generated by low-order Markov chains are already enough to get close to the capacity of sticky channels.
Interestingly, these numerical lower bounds strongly suggest (although they do not provide a rigorous proof) that the capacity of the geometric sticky channel is \emph{bounded away} from $0$ when the replication parameter $p\to 1$ (i.e., when the expected number of replications grows to infinity).

Although the numerical methods described above yield tight capacity lower bounds for the duplication and geometric sticky channels for fixed values of the replication parameter $p$ and provide some intuition about various properties of the capacity curve, they only provide a limited mathematical understanding of the behavior of these channels.
For example, these techniques give no rigorous insight about their behavior in limiting regimes, such as when $p\to 0$ or $p\to 1$.
Overall, it is unclear whether a numerical approach can get us closer to determining an \emph{exact} expression for the capacity of synchronization channels, which is arguably one of the main final goals of the study of such channels.

As a result, there is a natural need for \emph{analytical} capacity bounds for synchronization channels.
A few such lower bounds have already been presented in this section for the deletion channel, and more analytical bounds will be discussed in detail in later sections.
With respect to capacity lower bounds for sticky channels, both Drinea and Mitzenmacher~\cite{DM07} and Iyengar, Siegel, Wolf~\cite{ISW16} give analytical expressions for the rate achieved by an arbitrary order-$1$ Markov chain under both the duplication and geometric sticky channels (their approach can be generalized with some effort to Markov chains of higher orders).
The numerical bounds suggest that these analytical lower bounds are close to the true capacity, since we know that low-order Markov chains behave well under sticky channels.

Exploiting these analytical bounds, Iyengar, Siegel, and Wolf~\cite{ISW16} give a simple lower bound for the capacity of the geometric sticky channel with replication parameter $p$, which we denote by $C(\Geom_p)$, specialized for the $p\to 0$ regime.
More precisely, they show that
\begin{equation*}
    C(\Geom_p)\geq 1+p\log p+cp-O(p^2),
\end{equation*}
where $c\approx 0.8458$ is an explicit constant.
This lower bound is achieved by a uniform input distribution, and suggests that the geometric sticky channel behaves like a BSC for small replication parameter.
The same qualitative statement is known to hold true for the deletion channel with small deletion probability, as we shall see in Section~\ref{sec:lowdel}.

\paragraph{Lower bounds in the large alphabet setting}

All lower bounds we have seen thus far in this section have been presented for synchronization channels with binary input alphabet.
We remark that some of them can be generalized to synchronization channels with a $Q$-ary input alphabet.
The question of how the capacity of a $Q$-ary synchronization channel scales as $Q$ grows is natural, and it turns out to be more approachable than understanding the capacity of its binary counterpart.

For the capacity of the $Q$-ary deletion channel, which we denote by $C_Q(d)$, Mercier, Tarokh, and Labeau~\cite{MTL12} observe that lower bounds obtained by Diggavi and Grossglauser~\cite{DG06} imply that
\begin{equation}\label{eq:largeQdel}
    (1-d)\log Q-1\leq C_Q(d)\leq (1-d)\log Q.
\end{equation}
As a result, we conclude that
\begin{equation*}
    C_Q(d)\sim (1-d)\log Q
\end{equation*}
when $Q\to\infty$.
Note that the right-hand side of~\eqref{eq:largeQdel} is the capacity of the $Q$-ary erasure channel.
Therefore, for large $Q$ the $Q$-ary deletion channel behaves essentially like an erasure channel.
Remarkably, this turns out to also be true to a certain extent for large enough constant-sized alphabets against \emph{worst-case} deletions: Synchronization strings, introduced by Haeupler and Shahrasbi~\cite{HS17}, can be used to transform a code robust against worst-case erasures into one robust against worst-case deletions (with good parameters) with only a constant blow-up on the alphabet size. Such strings have found plenty of applications so far.

We note that Mercier, Tarokh, and Labeau~\cite{MTL12} derive large-alphabet capacity lower bounds for other types of synchronization channels too. As an example, for an \emph{arbitrary} sticky channel with replication distribution $R$ they show the bounds
\begin{equation}\label{eq:Qboundsdupl}
    \log Q-1\leq C_Q(R)\leq \log Q,
\end{equation}
where $C_Q(R)$ denotes the capacity of the $Q$-ary sticky channel under consideration.
From~\eqref{eq:Qboundsdupl}, we conclude that
\begin{equation*}
    C_Q(R)\sim \log Q
\end{equation*}
for \emph{every} replication distribution $R$.
The upper bound in~\eqref{eq:Qboundsdupl} is trivial. 
The lower bound is obtained by considering codewords $x\in[Q]^n$ for which no two consecutive symbols are the same, i.e., $x_i\neq x_{i+1}$ for all $i=1,\dots,n-1$.
Such codewords can be easily decoded with zero error probability by removing all consecutive duplicates of a symbol in the output.
Note that, by the discussion above, the bounds in~\eqref{eq:Qboundsdupl} can also be easily seen to apply to the \emph{zero-error} capacity of any sticky channel.

\section{General capacity upper bounds for synchronization channels}\label{sec:ub}

In this section, we present the main ideas and techniques behind several capacity upper bounds for many types of synchronization channels.
These include the deletion channel, sticky channels, and channels combining deletions and replications.
In the first part of this section, we will focus mostly on the deletion channel for the sake of clarity. Then, we move on to other types of synchronization channels.

Although non-trivial capacity lower bounds for the deletion channel have been known since the 1960's~\cite{Gal61,Zig69}, the first true non-trivial capacity upper bounds only appeared 40 years later~\cite{DMP07}.
Prior this, Ullman~\cite{Ull67} derived upper bounds for the \emph{zero-error} capacity of the deletion channel, and Dolgopolov~\cite{Dol90} obtained an upper bound on the rate achieved by a uniform input distribution under the deletion channel, assuming an unproven combinatorial conjecture.
However, none of these works yields capacity upper bounds in the i.i.d.\ deletions regime we are interested in.

In general, capacity upper bounds are obtained by revealing some extra side information about the input to the receiver (what is sometimes a \emph{genie-aided} argument).
This immediately leads to a modified channel with a higher capacity.
If the side information is chosen carefully, then the modified channel has a more approachable structure, and it is possible to determine or upper bound its capacity with known techniques.

\begin{figure}
	\centering
	\includegraphics[width=0.4\textwidth]{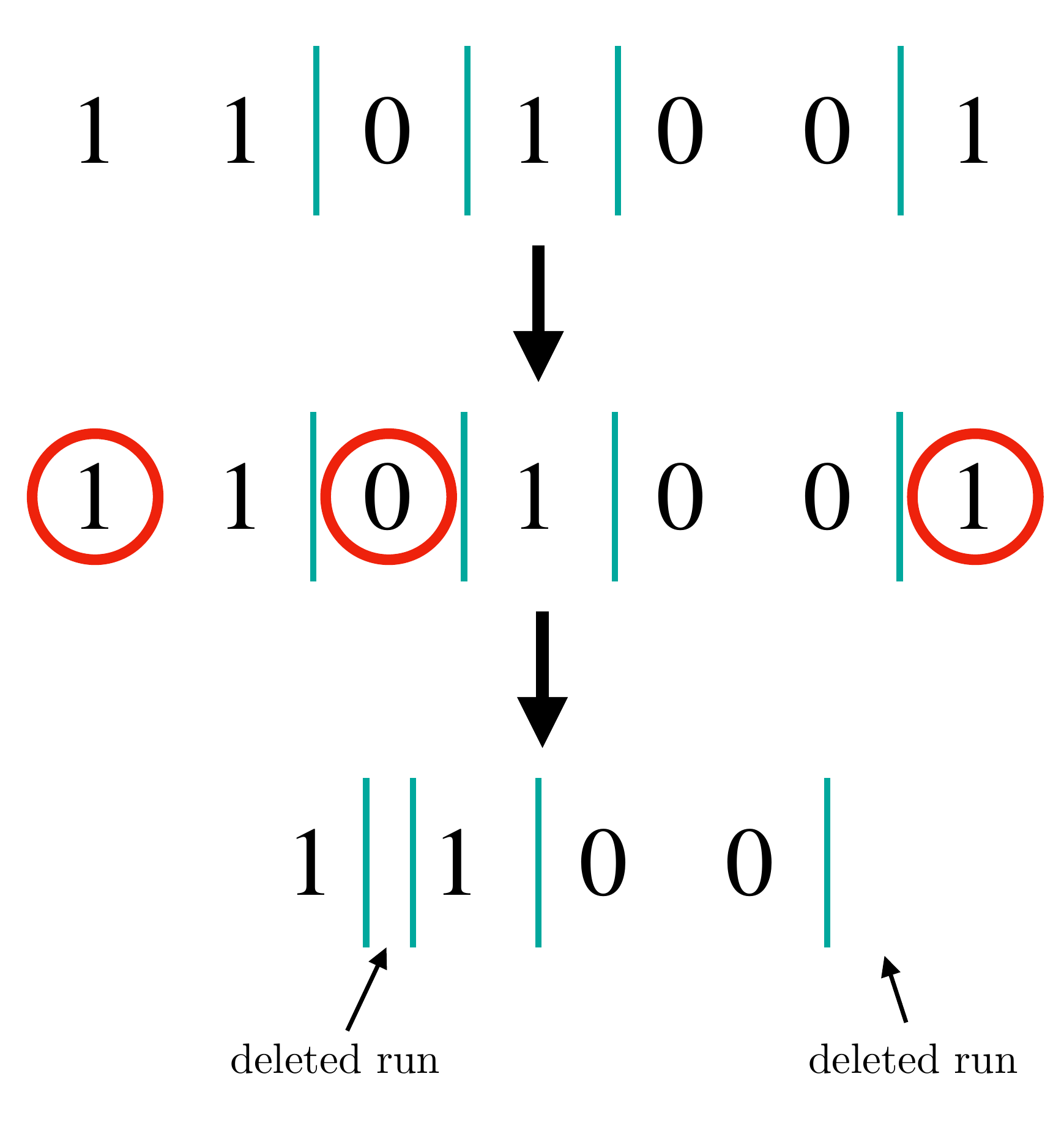}
	\caption{The deletion channel with undeletable markers. Circled bits are deleted.}
	\label{fig:markers}
\end{figure}

The first non-trivial capacity upper bound for the deletion channel (i.e., better than the trivial $1-d$ upper bound from the BEC) was derived by Diggavi, Mitzenmacher, and Pfister~\cite{DMP07}.
They consider a modified deletion channel that adds undeletable markers between input runs.
As a result, the receiver now knows which output bits come from each input run.
Figure~\ref{fig:markers} illustrates the modified channel with added markers.
The key property Diggavi, Mitzenmacher, and Pfister use is that the new channel is memoryless between different input runs.
Moreover, by the behavior of the deletion channel, the modified channel transforms input runs of length $u$ into output runs of length
\begin{equation}\label{eq:binchannel}
    V_u=\Bin_{u,1-d}.
\end{equation}
We call the channel $u\mapsto V_u=\Bin_{u,1-d}$ the \emph{binomial channel}.

Given what we have observed so far, we can relate the modified deletion channel to the binomial channel. 
Roughly speaking, since the modified channel is memoryless between input runs, we may essentially focus only on input distributions with i.i.d.\ run-lengths. 
Let $U$ be the run-length distribution of a given input distribution to the modified deletion channel.
Then, we expect to spend $\E[U]$ bits of the input to the modified deletion channel in each use of the binomial channel.
Therefore, our achievable rate for the modified deletion channel under run-length distribution $U$ should be
\begin{equation*}
    \frac{I(U;V_U)}{\E[U]},
\end{equation*}
where $V_U=\Bin_{U,1-d}$ is the output of the binomial channel.
Since the capacity of the deletion channel is trivially upper bounded by the capacity of the modified deletion channel, it follows that
\begin{equation}\label{eq:markerub}
    C(d)\leq \max_U\frac{I(U;V_U)}{\E[U]}=:\overline{\Ca}_\Bin(d).
\end{equation}
In words, the capacity of the deletion channel is upper bounded by the capacity \emph{per unit cost} of the binomial channel under the cost function $c(u)=u$ (recall Definition~\ref{def:capcost}), which we denote by $\overline{\Ca}_\Bin(d)$. Note that the binomial channel is a DMC, and hence we expect it to be much easier to handle than the deletion channel.

All it remains now is to obtain a good upper bound for the capacity per unit cost of the binomial channel, $\overline{\Ca}_\Bin(d)$.
In order to do this, Diggavi, Mitzenmacher, and Pfister make use of the following result of Abdel-Ghaffar~\cite{AG93}, which allows one to upper bound the capacity per unit cost of a DMC for arbitrary cost functions.
\begin{lem}[\cite{AG93}]\label{lem:AG}
    Consider a DMC $\Ch$ with discrete input alphabet $\cU\subseteq \mathbb{R}$, discrete output alphabet $\cV$, and conditional output distribution $V_u$.
    Then, for any distribution $Z$ over $\cV$ and any positive cost function $c$ we have
    \begin{equation}\label{eq:AGineq}
        \max_U \frac{I(U;V)}{\E[c(U)]}\leq\max_{u\in\cU} \frac{\KL(V_u||Z)}{c(u)}.
    \end{equation}
\end{lem}

If we wish to apply Lemma~\ref{lem:AG} to the binomial channel in view of~\eqref{eq:markerub}, we must deal with a maximization over an infinite input alphabet $\cU=\{1,2,\dots\}$.
This turns out to be difficult to work with directly.
If we could truncate the input alphabet of the binomial channel up to a finite threshold $m$, then we would reduce the right-hand side of~\eqref{eq:AGineq} to a finite maximization problem over $u\in\{1,2,\dots,m\}$. 
For a given choice of $Z$, this maximum could be easily computed.
However, this does not work directly, as truncating the input alphabet of the binomial channel decreases its capacity.
Instead, Diggavi, Mitzenmacher, and Pfister show how to carefully instantiate $Z$ in Lemma~\ref{lem:AG} so that (i) the infinite maximization problem is reduced to a finite one (which can be solved with computer assistance), and (ii) it leads to good upper bounds.
At a high level, $Z$ is constructed as follows: First, one determines the capacity-achieving \emph{output} distribution for the \emph{truncated} binomial channel (with some small threshold $m$) via the Blahut-Arimoto algorithm. This is a distribution with finite support. Then, a carefully chosen geometrically distributed tail is added to this finitely supported distribution to obtain $Z$.
This approach yields numerical capacity upper bounds for the deletion channel that improve upon the trivial $1-d$ upper bound for $d<0.9$.

Later, Fertonani and Duman~\cite{FD10} studied other types of side information, and improved upon the capacity upper bounds from~\cite{DMP07}.
At a high level, their strategy is to reduce the task of upper bounding (and even lower bounding) the capacity of the deletion channel to that of determining the capacity of a binary channel with \emph{fixed, finite} input length.
The latter task can be accomplished with computer assistance via the Blahut-Arimoto algorithm, provided that the input length considered is small.

\begin{figure}
	\centering
	\includegraphics[width=0.4\textwidth]{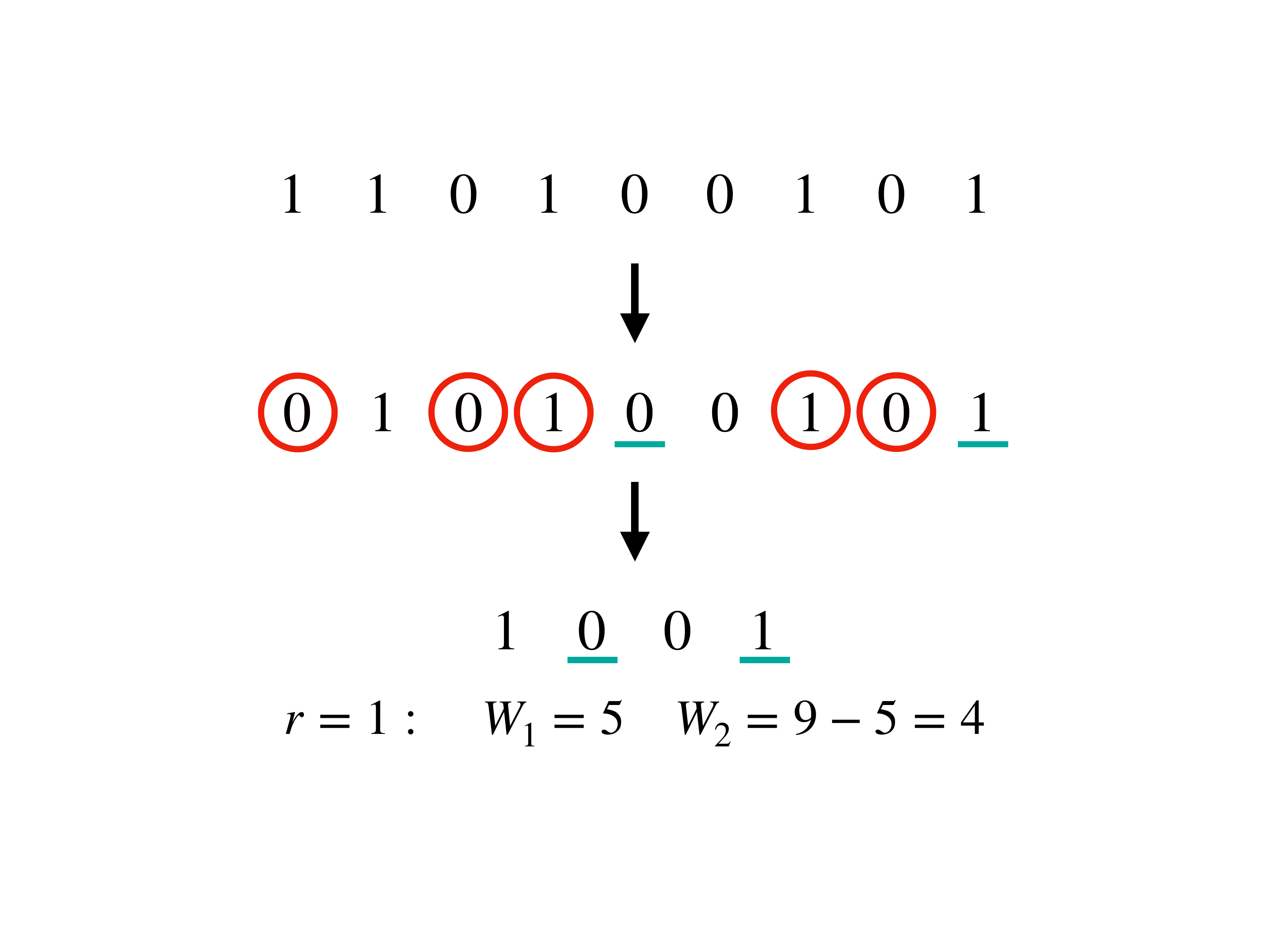}
	\caption{An example of the random process $W$ with $r=1$. Circled bits are deleted.}
	\label{fig:wprocess}
\end{figure}

The approach that leads to an improved capacity upper bound for the largest range of $d$ (actually, for $0.05<d<0.83$) has already been discussed in Section~\ref{sec:lb}.
To improve upon the upper bound from~\cite{DMP07} and the trivial upper bound $1-d$ for $d\geq 0.83$, Fertonani and Duman~\cite{FD10} consider another type of side information.
We proceed to describe the main idea.
They considered a genie that reveals to both sender and receiver a random process $W=(W_1,W_2,\dots)$ that works as follows: 
Fix some integer $r$,\footnote{As a small technicality, it must be assumed that the output length is a multiple of $r+1$. This does not affect the capacity of the channel.} and suppose $x$ and $y$ are the input and output, respectively, of the deletion channel. 
Then, $W_1$ denotes the index of the $(r+1)$-th received bit, $y_{r+1}$, in $x$. 
For $i>1$, $W_i$ denotes the difference between the indices in $x$ of $y_{(r+1)i}$ and $y_{(r+1)(i-1)}$. 
Note that revealing $W$ to sender and receiver can only increase the capacity, and splits the deletion channel into several independent channels, the $i$-th channel having $W_i$ input bits and $r+1$ output bits (see Figure~\ref{fig:wprocess} for an example with $r=1$).
With a little effort, one can write the capacity of the modified channel in terms of the distribution of $W_i$ and the capacity $C_{\ell,r}$ of the \emph{exact} deletion channel that receives $\ell$ bits and randomly deletes exactly $\ell-r$ bits. 
Two key observations then allow one to obtain a good upper bound on the capacity of the modified channel for all deletion probabilities $d$: 
First, $W_i$ follows a negative binomial distribution, and hence many quantities that appear in the bound simplify considerably. Second, for small input length $\ell$, it is computationally feasible to apply the Blahut-Arimoto algorithm to numerically approximate $C_{\ell,r}$.
It turns out this upper bound is also useful in the high deletion limiting regime $d\to 1$, as we shall see in Section~\ref{sec:highdel}.

As discussed in Section~\ref{sec:lb}, we can also obtain capacity upper bounds for the deletion channel directly from the well-known relationship (recall~\eqref{eq:sandwichcap} and~\eqref{eq:sandwichcapdel})
\begin{equation*}
    C_n(d)\geq C(d).
\end{equation*}
For small $n$, the quantity $C_n(d)$ can be computed using the Blahut-Arimoto algorithm.
However, this approach is prohibitive if $n$ is not very small.

The capacity upper bound obtained by Fertonani and Duman~\cite{FD10} is not convex for deletion probabilities $d\geq 0.65$.
Exploiting this, Rahmati and Duman~\cite{RD15} were able to improve on it for all $d\geq 0.65$.
This is done by proving a ``convexification'' result for the capacity of the deletion channel $C(d)$.
More precisely, Rahmati and Duman~\cite{RD15} proved that
\begin{equation}\label{eq:convexify}
    C(\lambda d^\star + 1-\lambda)\leq \lambda C(d^\star)
\end{equation}
for all $\lambda,d^\star\in[0,1]$.
In particular, setting $\lambda=\frac{1-d}{1-d^\star}$ for $d\geq d^\star$, we obtain
\begin{equation}\label{eq:convexifysimple}
    C(d)\leq \frac{1-d}{1-d^\star}C(d^\star)
\end{equation}
for any $d\geq d^\star$.
As a result, we conclude that any upper bound for $C(d^\star)$ can be extended linearly to every $d\geq d^\star$.
Considering~\eqref{eq:convexifysimple} with $d^\star=0.65$ and replacing $C(0.65)$ by the best capacity upper bound for this deletion probability immediately improves upon the capacity upper bound of Fertonani and Duman~\cite{FD10}.

In fact, Rahmati and Duman proved a stronger result relating $C(\lambda d_1+(1-\lambda)d_2)$ with $\lambda C(d_1)$ and $(1-\lambda)C(d_2)$ for arbitrary $\lambda,d_1,d_2\in[0,1]$.
However, the special case presented in~\eqref{eq:convexify} (which corresponds to $d_2=1$) is the one that currently leads to improved capacity upper bounds.

\begin{figure}
	\centering
	\includegraphics[width=0.5\textwidth]{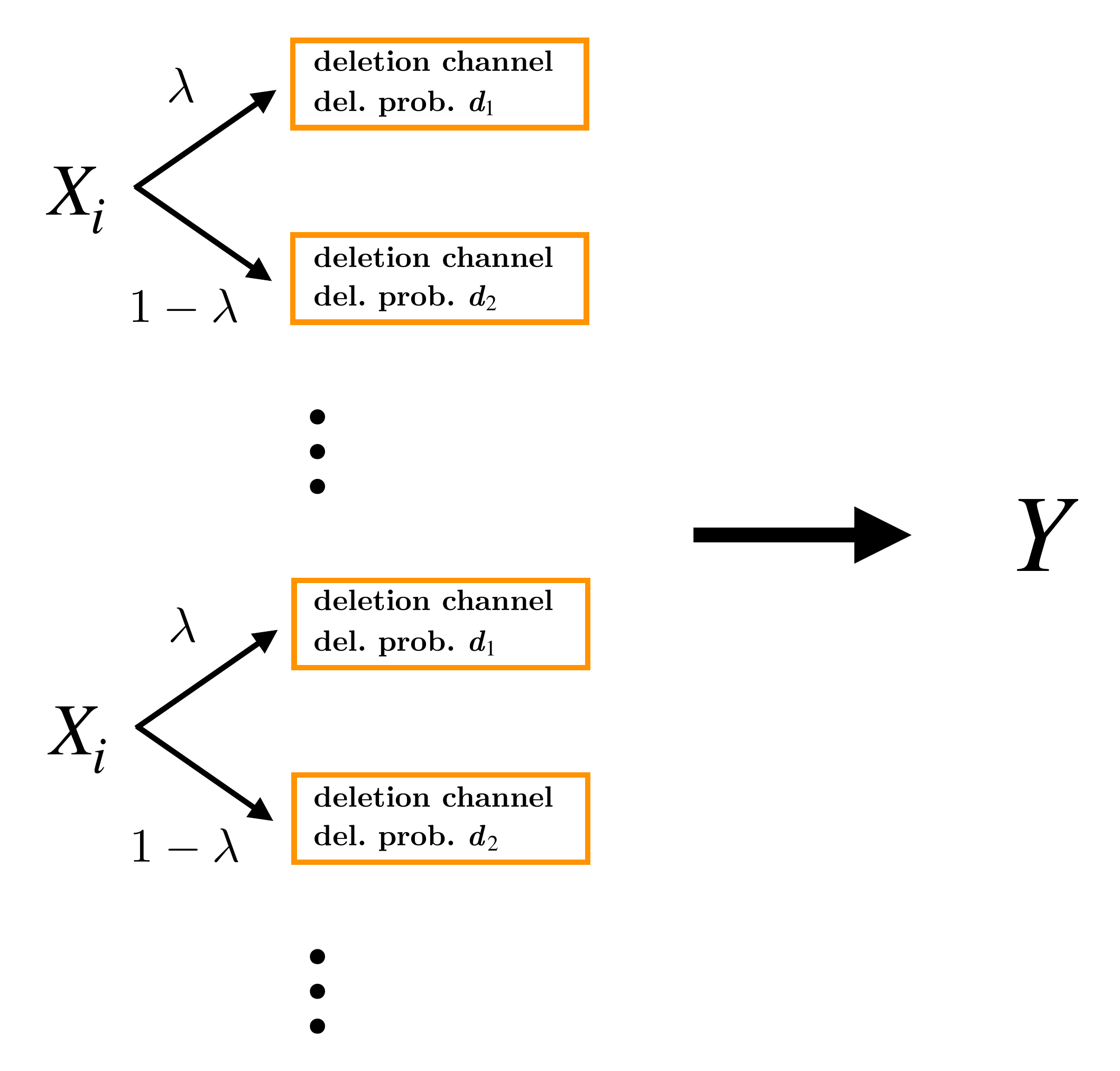}
	\caption{A fragmented deletion channel. This channel is equivalent to a deletion channel with deletion probability $\lambda d_1+(1-\lambda)d_2$.}
	\label{fig:channelfrag}
\end{figure}

The inequality in~\eqref{eq:convexify} is proved via what Rahmati and Duman call \emph{channel fragmentation}.
Consider the following deletion process: There are two independent deletion channels $\Ch_1$ and $\Ch_2$ with deletion probabilities $d_1$ and $d_2$, respectively. Each input bit is sent through $\Ch_1$ with probability $\lambda$ or through $\Ch_2$ with probability $1-\lambda$.
Figure~\ref{fig:channelfrag} illustrates the behavior of this channel.
Of course, this channel is equivalent to a deletion channel with deletion probability $\lambda d_1+(1-\lambda)d_2$.
Nevertheless, this ``fragmented'' view of the deletion channel naturally leads to good capacity upper bounds.

For the case $d_1=d^\star$ and $d_2=1$, Rahmati and Duman obtain~\eqref{eq:convexify} via a simple argument.
Suppose $X$ is the $n$-bit input to the fragmented deletion channel described in the previous paragraph with $d_1=d^\star$ and $d_2$=1, and denote the corresponding output by $Y$. Furthermore, let $X^{[1]}$ denote the input bits that go through the channel $\Ch_1$, with $Y^{[1]}=Y$ being the corresponding output.
Since $X\rightarrow X^{[1]}\rightarrow Y^{[1]}=Y$ is a Markov chain, by the data processing inequality we have
\begin{equation}\label{eq:dataproc}
    I(X;Y)\leq I(X^{[1]};Y^{[1]}).
\end{equation}
Observe that $|X^{[1]}|$ follows a $\Bin_{n,\lambda}$ distribution, and so $|X^{[1]}|$ is close to its expected value $\lambda n$ with high probability.
If $|X^{[1]}|=\lambda n$, then we could immediately upper bound $I(X^{[1]};Y^{[1]})$ as (recall~\eqref{eq:finitelength})
\begin{equation*}
    I(X^{[1]};Y^{[1]})\leq \lambda n\cdot \Ca_{\lambda n}(\Ch_1).
\end{equation*}
While the assumption above is not true, the fact that $|X^{[1]}|$ is nevertheless close to $\lambda n$ with high probability means leads to the inequality
\begin{equation}\label{eq:noassump}
    I(X^{[1]};Y^{[1]})\leq \lambda n\cdot \Ca_{\lambda n}(\Ch_1)+o(n).
\end{equation}
Therefore,~\eqref{eq:convexify} follows by combining~\eqref{eq:dataproc} and~\eqref{eq:noassump}, dividing everything by $n$, and taking the limit $n\to\infty$.

A similar ``channel fragmentation'' techniques was also used by Rahmati and Duman~\cite{RD13b} to derive capacity upper bounds for the $2Q$-ary deletion channel directly from any capacity upper bound for the binary deletion channel.
This approach is useful because adapting the strategies discussed above to the $2Q$-ary alphabet setting is not feasible from a computational perspective.
This is because the complexity of the Blahut-Arimoto-type algorithms used in the works above grows quickly with the alphabet size.

Note that all capacity upper bounds discussed so far are obtained by first reducing the (hard) optimization problem of determining the capacity of the deletion channel to a finite optimization problem which is amenable to off-the-shelf numerical methods.
Analogously to what was discussed in Section~\ref{sec:lb}, although such a strategy results in very good numerical upper bounds on the capacity of the deletion channel for fixed deletion probabilities and aids our intuition, it provides limited rigorous insight into the actual behavior of the capacity curve.
This suggests that we may need a new approach if we hope to get a more complete understanding of the capacity of synchronization channels.

Motivated by this, Cheraghchi~\cite{Che19} focused on obtaining \emph{analytical} capacity upper bounds for the deletion channel (among other repeat channels) that require as little computer assistance to be derived as possible.
Nevertheless, his capacity upper bounds also improve upon previous ones for $d\leq 0.02$.
Besides what has been said above, analytical capacity bounds open the door to a mathematically rigorous approach to synchronization channels.
In particular, by analyzing an analytical capacity bound, one may hope to prove good asymptotic results about the capacity curve, or obtain sharp closed form bounds on the capacity.
Moreover, this approach leaves the door open for a derivation of the \emph{exact} capacity of a synchronization channel.

The strongest upper bounds in~\cite{Che19} are given by maximums of concave smooth functions (given by sums of explicit exponentially decaying sequences) over $[0,1]$.
As a result, this maximization problem can be solve to the desired accuracy efficiently.
Moreover, the amount of computer assistance required can be reduced further by replacing the concave smooth functions above (given in terms of infinite sums) by sharp upper bounds in terms of elementary and special functions.
Notably, in some cases the resulting maximization problem can be solved analytically, leading to non-trivial capacity upper bounds with \emph{human-readable} proofs.
A particular example of this is the capacity upper bound
\begin{equation}\label{eq:fullyexplicit}
    C(1/2)\leq \frac{\log \varphi}{2},
\end{equation}
where $\varphi=\frac{1+\sqrt{5}}{2}$ is the golden ratio.

The starting point for Cheraghchi~\cite{Che19} is, similarly to previous works, the addition of carefully chosen side information to obtain a suitable upper bound.
For the special case of the deletion channel, he obtains
\begin{equation}\label{eq:ChcapUB1}
    C(d)\leq (1-d)\cdot\sup_{\mu \geq 0}\frac{\Ca_\mu(\Bin_{1-d})}{1+\mu},
\end{equation}
where $\Ca_\mu(\Bin_{1-d})$ denotes the capacity of the binomial channel with \emph{mean constraint $\mu$} and success probability $1-d$.
This is a channel that behaves like the binomial channel already discussed above in the context of the upper bounds obtained by Diggavi, Mitzenmacher, and Pfister~\cite{DMP07}, but which only accepts input distributions $U$ such that $\E[U]=\mu$.

Note that~\eqref{eq:ChcapUB1} looks similar to~\eqref{eq:markerub} from~\cite{DMP07} where $C(d)$ is upper bounded by the capacity per unit cost of the binomial channel. This is not a coincidence, and indeed the right-hand side of~\eqref{eq:ChcapUB1} can be expressed as a capacity per unit cost.
One may wonder whether it would be feasible to obtain good analytical capacity upper bounds for the deletion channel by considering~\eqref{eq:markerub} and applying Lemma~\ref{lem:AG} with a carefully chosen $Z$ given by an analytical expression.
This turns out to be very complicated, even for simple cost functions like $c(u)=u$.
The change of perspective to mean-limited channels in~\eqref{eq:ChcapUB1} is performed to enable a similar approach to actually work.

The second step in~\cite{Che19} is to establish a good way of obtaining analytical upper bounds for $\Ca_\mu(\Bin_{1-d})$, i.e., the capacity of a \emph{mean-limited} channel.
This is accomplished by casting the problem of determining the capacity as a convex program, and employing techniques from convex duality and the Karush-Kuhn-Tucker (KKT) conditions. One then obtains the following result.
\begin{lem}[\protect{\cite{Che19}, specialized}]\label{lem:ubmeanlimited}
    Suppose there is a distribution $Z$ over $\cV$ and constants $a,b\in\mathbb{R}$ such that
    \begin{equation*}
        \KL(V_u||Z)\leq a\E[V_u]+b
    \end{equation*}
    for every $u\in\cU$, where recall $V_u=\Bin_{u,1-d}$.
    Then, it follows that
    \begin{equation*}
        \Ca_\mu(\Bin_{1-d})\leq a\mu+b
    \end{equation*}
    for every $\mu\geq 0$.
    Moreover, we have $\Ca_\mu(\Bin_{1-d})= a\mu+b$ if and only if:
    \begin{enumerate}
        \item $Z$ is a valid output distribution of the mean-limited binomial channel with some associated input distribution $U'$;
        \item We have $\KL(V_u||Z)=a\E[V_u]+b$ for all $u\in\supp(U')$.
    \end{enumerate}
\end{lem}

We remark that Lemma~\ref{lem:ubmeanlimited} gives both a way of computing capacity upper bounds for mean limited channels and also a way of checking whether one has obtained an exact expression for the channel capacity.
Results of this type are well-known in the information theory literature, and have been used to study the capacity of several different channels (e.g., see~\cite{Smi71,Sha90,AFTS01,Mar07,LM09}).

Coupling~\eqref{eq:ChcapUB1} with Lemma~\ref{lem:ubmeanlimited} and careful choices of $Z$, Cheraghchi was able to derive several analytical capacity upper bounds.
An example of a good choice of $Z$ obtained through a convexity-based argument is the so-called \emph{inverse binomial distribution} $\InvBin_q$ defined as
\begin{equation*}
    \InvBin_q(y)=y_0\binom{y/p}{y}q^y\exp(-y h(p)/p),\quad y=0,1,2,\dots,
\end{equation*}
where $p=1-d$, $y_0$ is the normalization factor, and $q\in (0,1)$ is a free constant.
The inverse binomial distribution leads to good analytical capacity upper bounds for the deletion channel.
Furthermore, it is possible to sharply bound it in terms of both elementary functions and standard special functions such as the Lerch transcendent.
A particularly notable case is when $d=1/2$. Then, the inverse binomial distribution becomes a negative binomial distribution.
In this case, the resulting maximization problem in the capacity upper bound can be solved \emph{without} computer assistance, leading to the fully explicit bound in~\eqref{eq:fullyexplicit}.

Cheraghchi constructed various candidate distributions $Z$ not only for the deletion channel, but also for the Poisson-repeat channel. 
However, we note that none of the distributions considered in~\cite{Che19} satisfy either of the two conditions required for optimality in Lemma~\ref{lem:ubmeanlimited}.

Li~\cite{Li18} also uses optimization techniques to derive maximum likelihood upper bounds for information stable channels, and, among other things, specializes his approach for the deletion channel.
He reduces upper bounding the capacity of the deletion channel to solving a certain combinatorial problem over $\bits^n$.
Namely, this problem asks to, given an output string $y$ from the deletion channel, to find $x\in\bits^n$ such that the number of deletion patterns that transform $x$ into $y$ is maximized.
Although this combinatorial problem is hard to tackle directly, Li considers more efficient ways of approximating its solution.

\begin{figure}
	\centering
	\includegraphics[width=0.65\textwidth]{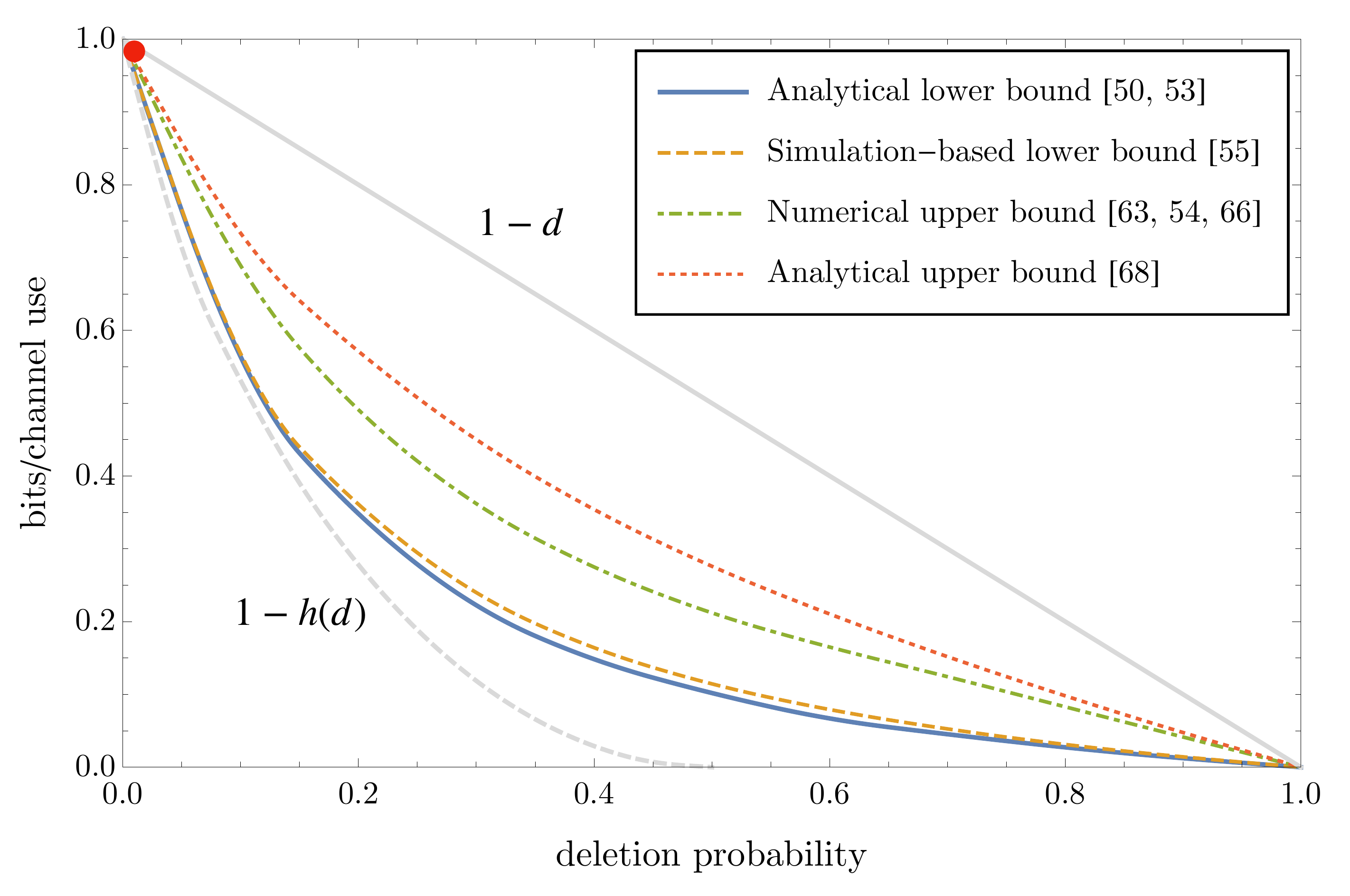}
	\caption{Capacity bounds for the deletion channel. The light gray lines depict the capacities of the BEC and the BSC. The circle marks where the analytical capacity upper bound from~\cite{Che19} beats the numerical capacity upper 
	bound from~\cite{DMP07,FD10,RD15}.}
	\label{fig:capdel}
\end{figure}

Figure~\ref{fig:capdel} collects the best analytical and simulation-based lower bounds along with the best analytical and numerical upper bounds on the capacity of the deletion channel.

\paragraph{Capacity upper bounds for sticky channels}

As already discussed in Section~\ref{sec:lb}, the capacity of a sticky channel $\Ch$ equals the capacity per unit cost of a certain DMC $\Ch'$ over the positive integers with cost function $c(x)=x$ (recall Definition~\ref{def:capcost} and~\eqref{eq:equivcapunitcost}).
In particular, when $\Ch$ is the duplication channel, then $\Ch'$ maps $x\in\{1,2,\dots\}$ to $Y_x=x+\Bin_{x,p}$ for some replication parameter $p$.
Moreover, if $\Ch$ is the geometric sticky channel, then $\Ch'$ maps $x\in\{1,2,\dots\}$ to $Y_x=x+\NB_{x,p}$, where recall $\NB_{x,p}$ denotes a negative binomial distribution with $x$ failures and success probability $p$.

This observation suggests a clear approach towards obtaining numerical capacity upper bounds for sticky channels. Namely, similarly to what was done by Diggavi, Mitzenmacher, and Pfister~\cite{DMP07} for the deletion channel, one can exploit Lemma~\ref{lem:AG} with a careful choice of $Z$ to derive upper bounds on the capacity per unit cost of DMC's.
This was the strategy originally undertaken by Mitzenmacher~\cite{Mit08}.
As was done in~\cite{DMP07}, Mitzenmacher constructs $Z$ by combining a capacity-achieving input distribution for the truncated DMC $\Ch'$ with a suitable geometric tail.
This leads to tight numerical capacity upper bounds for the duplication channel.
However, he was not able to derive capacity upper bounds for the geometric sticky channel.

Later, Mercier, Tarokh, and Labeau~\cite{MTL12} used the same high-level strategy from the previous paragraph to derive tight capacity upper bounds for the geometric sticky channel (and other similar channels).
However, their low-level approach differs from that of Mitzenmacher~\cite{Mit08}, as they construct a distribution $Z$ using different weights for small inputs (instead of using the output distribution associated to the capacity-achieving input distribution for the truncated DMC) along with a geometric tail.

\begin{figure}
	\centering
	\includegraphics[width=0.65\textwidth]{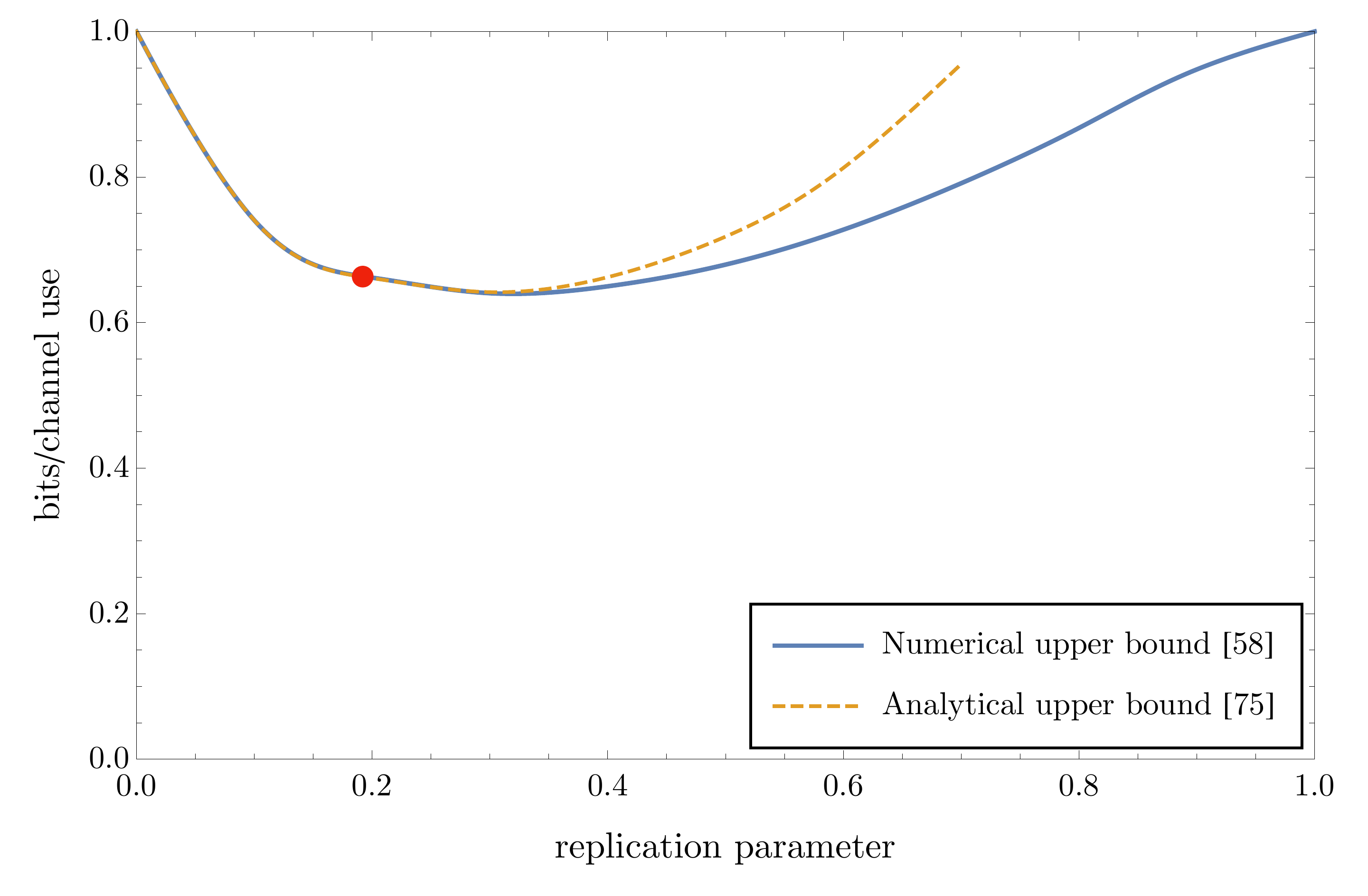}
	\caption{Tight capacity upper bounds for the duplication channel. The circle indicates where the analytical upper bound from~\cite{CR19b} beats the numerical upper bound from~\cite{Mit08}.}
	\label{fig:capdupl}
\end{figure}

\begin{figure}
	\centering
	\includegraphics[width=0.65\textwidth]{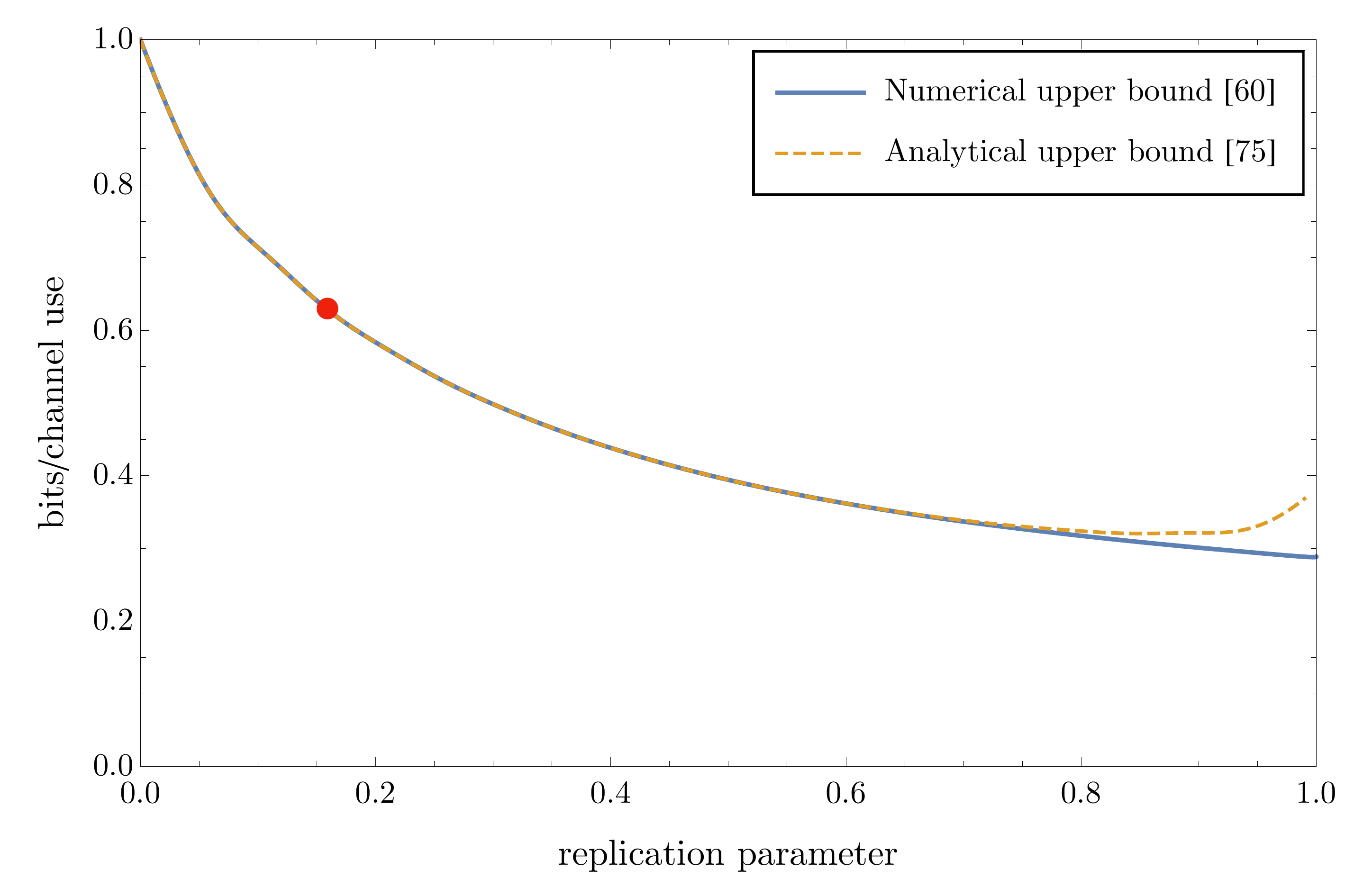}
	\caption{Tight capacity upper bounds for the geometric sticky channel. The circle indicates where the analytical upper bound from~\cite{CR19b} beats the numerical upper bound from~\cite{MTL12}.}
	\label{fig:capgeom}
\end{figure}

Cheraghchi and Ribeiro~\cite{CR19b}, building on techniques from~\cite{Che19}, derived tight \emph{analytical} capacity upper bounds for the duplication and geometric sticky channels.
As before, these bounds are maximums of concave smooth functions over $[0,1]$, and hence can be computed efficiently to any desired accuracy. Moreover, besides being analytical, these upper bounds improve upon on the previous numerical bounds~\cite{Mit08,MTL12} for some choices of the replication parameter $p$.
Similarly to the connection between the capacity of sticky channels and the capacity per unit cost of certain DMC's, we may also write
\begin{equation*}
    \Ca(\Ch)=\lambda \sup_{\mu\geq \lambda}\frac{\Ca_\mu(\Ch')}{\mu},
\end{equation*}
where $\Ch$ is a sticky channel with replication distribution $R$, $\lambda=\E[R]$, and $\Ca_\mu(\Ch')$ is the capacity of a certain DMC $\Ch'$ over the integers with mean constraint $\mu$.
Therefore, it suffices to apply Lemma~\ref{lem:ubmeanlimited} with a careful choice of the candidate distribution $Z$ to obtain good capacity upper bounds for sticky channels.
Cheraghchi and Ribeiro are able to construct a family of distributions $Z$ such that
\begin{equation}\label{eq:zerogap}
    \KL(V_u||Z)=a\E[V_u]+b
\end{equation}
for \emph{every} $u$, where $V_u$ denotes the output distribution of the DMC $\Ch'$ on input $u$.
Intuitively, this is one of the reasons for the high quality of the resulting analytical capacity upper bounds.
Observe that~\eqref{eq:zerogap} means Condition $2$ for the optimality of $Z$ in Lemma~\ref{lem:ubmeanlimited} is automatically satisfied.
However, they were not able to find a $Z$ in the family satisfying Condition 1 for optimality.
If such a $Z$ is found (either directly or by adapting the techniques), then one obtains an \emph{exact} expression for the capacity of the underlying sticky channel.
Figures~\ref{fig:capdupl} and~\ref{fig:capgeom} plot the best capacity upper bounds for the duplication and geometric sticky channels.


\paragraph{Capacity upper bounds for channels with replications and deletions}

Some works also study the capacity of channels combining deletions and replications.
Mercier, Tarokh, and Labeau~\cite{MTL12} adapt their approach towards obtaining capacity upper bounds for sticky channels to channels with deletions and replications by considering extra side information such as undeletable markers, as in~\cite{DMP07}.
Cheraghchi~\cite{Che19} obtained analytical capacity upper bounds for the Poisson-repeat channel using the high-level approach discussed above for the deletion channel (but using a very different low-level analysis).
Techniques developed in~\cite{Che19} were later adapted to yield a better understanding of the capacity of the discrete-time Poisson channel~\cite{CR19a}, which is an important channel in optical communications.
Finally, Cheraghchi and Ribeiro~\cite{CR19b} studied the \emph{geometric deletion channel}, which independently replicates bits according to a $\Geom_p$ distribution.
They obtain analytical capacity upper bounds for this channel, and notably give a fully analytical proof that the capacity of this channel is at most $0.73$ bits/channel use (hence \emph{bounded away from $1$}) in the limit $p\to 1$ (which corresponds to the expected number of replications growing to infinity).\footnote{We note this is not obvious for repeat channels with infinitely supported replication distributions. In fact, the capacity of the Poisson-repeat channel converges to $1$ when the expected number of replications grows to infinity~\cite{CR19b}.}
Thus far, this is the only non-trivial capacity upper bound for any repeat channel in the ``many replications'' regime.


\section{The deletion channel capacity in limiting regimes}\label{sec:limit}

Since determining the deletion channel capacity appears to be very difficult, significant attention has been directed at characterizing this capacity in the regimes where the deletion probability $d$ approaches either $0$ or $1$. The case where $d\to 0$ is solved (up to lower order terms), while the gap between upper and lower bounds in the $d\to 1$ setting is still considerable. This section aims to provide an account of the results and techniques used in these limiting regimes.

\subsection{Low deletion probability}\label{sec:lowdel}

In this section, we discuss of the concurrent efforts that succeeded in determining the high-order terms of the capacity of the deletion channel in the limit $d\to 0$~\cite{KMS10,KM13}. Notably, both works use radically different approaches.

Our starting point is the basic lower bound on $C(d)$~\cite{Gal61,Zig69,DG06}, 
\begin{equation*}
    C(d)\geq 1-h(d).
\end{equation*}
As discussed in Section~\ref{sec:lb}, the result is obtained by considering a uniform input distribution.
This bound is quite bad for $d$ not too small (in particular, for $d$ close to $1/2$). However, as we shall see, this bound is essentially optimal for small $d$. Intuitively, this seems to make sense. In fact, the uniform input distribution is trivially capacity-achieving at $d=0$. Therefore, one expects that it should also behave quite well for small deletion probability. Kalai, Mitzenmacher, and Sudan~\cite{KMS10} gave a rigorous basis to this intuition using an interesting counting argument, and Kanoria and Montanari~\cite{KM13} provided a tighter asymptotic analysis, leading to the result below.
\begin{thm}[\cite{KMS10,KM13}]\label{thm:lowdel}
    For every constant $\eps>0$, it holds that
    \begin{equation*}
        C(d) = 1-h(d)+c_1d+ c_2 d^2 + O(d^{3-\eps}),
    \end{equation*}
    where $c_1\approx 1.1542$ and $c_2\approx 1.6781$. This rate is achieved by:
   \begin{itemize}
       \item The uniform input distribution, to within $O(d^{2-\eps})$~\cite{KM10};
       \item An input distribution with i.i.d.\ runs, to within $O(d^{3-\eps})$.
   \end{itemize}
\end{thm}
Observe that Theorem~\ref{thm:lowdel} shows that the deletion channel with small deletion probability $d$ behaves almost like a BSC with error probability $d$.

Kalai, Mitzenmacher, and Sudan~\cite{KMS10} determined the weaker expansion
\begin{equation*}
    C(d)=1-h(d)+O(d).
\end{equation*}
Their approach is based on a different way of looking at capacity upper bounds: 
Suppose that for the deletion channel there is a decoder that, with decent probability, is able to recover both the input \emph{and} the positions of the deletions applied by the channel from the channel output. 
Then, this leads to a good capacity upper bound. In fact, consider, for the sake of clarity, the \emph{exact} deletion channel $\mathsf{Del}_{n,q}$ that receives $n$ input bits and deletes exactly $q$ of them at random\footnote{For large input blocklength, it is easy to translate results between the exact deletion channel with $q=\lceil dn\rceil$ deletions and the usual deletion channel with deletion probability $d$ via standard concentration arguments.}.
This exact channel has also been considered in other works on deletion capacity bounds discussed before.
For an input $x\in\bits^n$ to $\Del_{n,q}$, the output $y=\in\bits^{n-q}$ can be written as $y=x_p$ for a deletion pattern $p\in\bits^n$ with $p_i=1$ if and only if $x_i$ is deleted. Suppose there exist a code $\cC\subseteq\bits^n$ and a deterministic decoder $\Dec:\bits^{n-q}\to \bits^n\times \bits^n$ such that
\begin{equation}\label{eq:conddec}
    \Pr_{X\leftarrow \cC, P\leftarrow\cP}[\Dec(X_P)=(X,P)]\geq \gamma.
\end{equation}
The inequality in~\eqref{eq:conddec} means there are at least $\gamma\cdot|\cC|\cdot |\cP|=\gamma\cdot|\cC|\cdot \binom{n}{q}$ pairs $(x,p)$ such that $\Dec(x_p)=(x,p)$. On the other hand, each such good pair is mapped to a different channel output in $\bits^{n-q}$. This immediately leads to the bound
\begin{equation}\label{eq:basicubkms}
    \gamma\cdot|\cC|\cdot \binom{n}{q}\leq 2^{n-q},
\end{equation}
which implies, by setting $q=\lceil dn\rceil$ and using the standard equality
\begin{equation*}
    \log\binom{n}{q}=n\cdot h(d)(1+o(1)),
\end{equation*}
the following upper bound on the rate of $\cC$:
\begin{equation*}
    \frac{\log|\cC|}{n}\leq 1-\frac{q}{n}-\frac{1}{n}\log\binom{n}{q}-\frac{1}{n}\log \gamma\approx 1-d-h(d)-\frac{1}{n}\log \gamma.
\end{equation*}
Therefore, if we come up with a decoder whose success probability $\gamma$ is not too small, then we obtain the desired $1-h(d)+O(d)$ upper bound.

Remarkably, when applied to simpler channels such as the BSC and the BEC, this approach easily yields the correct upper bounds on their capacities. In both channels, any decoder that succeeds in recovering the input $x$ from the output $x_p$ (where in the error pattern one now replaces deletions with bit-flips or erasures) automatically recovers $p$ correctly too. This means we can set the success probability $\gamma=1-o(1)$. With respect to the BSC, there are $2^n$ possible output sequences, and so one obtains, analogously to~\eqref{eq:basicubkms},
\begin{equation*}
    (1-o(1))\cdot|\cC|\cdot \binom{n}{q}\leq 2^n.
\end{equation*}
Setting $q=\lceil dn\rceil$, this implies the correct capacity upper bound
\begin{equation*}
    \frac{\log|\cC|}{n}\leq 1-h(d)+o(d)
\end{equation*}
for the BSC with error probability $d$. For the BEC, the only difference is that the number of possible output sequences is $2^{n-q}\cdot \binom{n}{q}$, which leads to the correct bound (via~\eqref{eq:basicubkms} with $q=\lceil dn\rceil$)
\begin{equation*}
    \frac{\log|\cC|}{n}\leq 1-d+o(1)
\end{equation*}
for the BEC with erasure probability $d$.

Going back to the deletion channel, in view of~\eqref{eq:conddec} and~\eqref{eq:basicubkms} all that is required now is to show that all sufficiently large codes $\cC$ that can be decoded with high probability also have a decoder $\Dec$ that successfully decodes the channel output \emph{and} recovers the correct deletion pattern with decent probability $\gamma$. However, this is not as simple as for the BSC and BEC, since many deletion patterns may yield the same output for a fixed input string, and all of them are equally likely. Given this, the natural strategy (followed in~\cite{KMS10}) is to show that for most codewords of $\cC$ few deletion patterns lead to the same output, provided the deletion probability is small. This means the most obvious choice of $\Dec$ works: Fix $\cC\subseteq \bits^n$ that can be decoded with probability $1-o(1)$ via some decoder $\Dec':\bits^{n-q}\to\bits^n$, and let $Y=X_P$ for $X\leftarrow\cC$ and $P\leftarrow\cP$. Consider $\Dec:\bits^{n-q}\to\bits^n\times\bits^n$ that on input $Y$ computes $\hat{X}=\Dec'(Y)$, and then finds the first pattern (say, in lexicographic order) of $q$ deletions $\hat{P}\in\bits^n$ such that $\hat{X}_{\hat{P}}=Y$. If such pattern exists, the output of $\Dec$ is $\Dec(Y)=(\hat{X},\hat{P})$. Provided that for most choices of $X$ there are few patterns $p$ such that $X_p=Y$, then, if $\hat{X}=X$ (which happens with high probability), it follows that $\hat{P}=P$ with decent probability, since all possible deletion patterns are equally likely.

Using a different approach, Kanoria and Montanari~\cite{KM13} determined the remaining low-order terms of the capacity in Theorem~\ref{thm:lowdel}. We remark that this requires more refined upper \emph{and} lower bounds on the capacity: The basic lower bound using a uniform input distribution is not strong enough. Nevertheless, similarly to what was mentioned before, given that a uniform input distribution is trivially capacity-achieving at $d=0$, it makes sense that the true capacity-achieving distribution for small $d$ is a slight perturbation of the uniform input distribution.

An initially non-rigorous argument assuming that consecutive runs experience very few deletions (due to the small deletion probability) suggests that a certain input distribution with i.i.d.\ runs and run-length distribution $P$ satisfying $\Pr[P=\ell]\approx 2^{-\ell}$ should achieve nearly optimal rate in the small deletion probability regime (observe that a run in the uniform input distribution has length $\ell$ with probability exactly $2^{-\ell}$). The achievability part of Theorem~\ref{thm:lowdel} is obtained by directly estimating the rate of this explicit input distribution.
We note that an alternative proof of the achievability part of Theorem~\ref{thm:lowdel} was given by Iyengar, Siegel, and Wolf~\cite{ISW16}.
At a high level, the proof of the converse follows by showing that (i) the capacity is achieved by a stationary ergodic process, (ii) every such process with high entropy rate (necessary to improve on the rate achieved of the explicit input distribution with i.i.d.\ runs) must have run-length distribution very close in statistical distance to that of the uniform input distribution, and (iii) one may indeed consider a modified deletion process where deletions are forced to be far apart (a bit more precisely, there can be at most two deletions per input run) with only an $O(d^3)$ error in the true capacity.

To conclude this section, we note that Ramezani and Ardakani~\cite{RA13} applied techniques similar to those used in~\cite{KM13} to the duplication channel with small duplication probabilty. In that setting, they also conclude that the duplication channel behaves close to a BSC.

\subsection{High deletion probability}\label{sec:highdel}

As we have seen before, the deletion capacity in the small deletion probability regime is very well understood (up to low-order terms). Given this, it is natural to look at the opposite limiting regime and study the deletion capacity when the deletion probability is close to $1$ instead. There, our knowledge still has some large gaps.

We start with the basic connection between the deletion channel and the BEC. As discussed before, the simple observation that a deletion is harder to handle than an erasure leads to the upper bound
\begin{equation*}
    C(d)\leq 1-d,
\end{equation*}
where $1-d$ is the capacity of the BEC with erasure probability $d$. A natural question arises: We know that the deletion channel is a harder channel than the BEC, but is it \emph{significantly} harder? A bit more precisely, is it true that $C(d)=o(1-d)$ when $d\to 1$?

Mitzenmacher and Drinea~\cite{MD06} showed that the answer to this question is negative: The deletion channel and the BEC are \emph{always} of comparable difficulty, in the sense that there exists a constant $c>0$ such that $C(d)\geq c(1-d)$ for all $d$.
\begin{thm}\label{thm:poissonlb}
    For every $d$, we have $0.1185(1-d)\leq C(d)\leq 1-d$.
\end{thm}
The ideas behind this lower bound are very well covered in Mitzenmacher's survey~\cite{Mit09}. 
For completeness, we present an outline of their proof. At the basis of their approach lies the following observation: 
Suppose we generate a run of length distributed according to $\Poi_{\frac{\lambda}{1-d}}$, and send it through a deletion channel with deletion probability $d$. 
Then, the length of the output is distributed according to $\Poi_{\lambda}$. 
This suggests the following reduction from the deletion channel to the Poisson-repeat channel: 
Sending a bit-string $x\in\bits^n$ through the Poisson-repeat channel with mean $\lambda$ is equivalent to first replicating each bit of $x$ independently according to $\Poi_{\frac{\lambda}{1-d}}$, and then sending the resulting string $x'$ through a deletion channel with deletion probability $d$. 
Suppose we have a codebook $\cC_\Poi$ of rate $R$ for the Poisson-repeat channel where each bit is replicated $\Poi_{\frac{\lambda}{1-d}}$ times. 
Then, ignoring some technicalities\footnote{In particular, the failure probability of the decoder for $\cC_\Poi$ must be at most $1/n^c$ for some $c>1/2$, where $n$ is the blocklength.} and recalling that each bit is replicated $\frac{\lambda}{1-d}$ times in expectation, applying the replication process to every codeword in $\cC_\Poi$ leads to a good codebook for the deletion channel with rate
\begin{equation*}
    \frac{(1-d) R}{\lambda}.
\end{equation*}
Since the reduction above works for every $d$ and $\lambda$, we have the inequality
\begin{equation}\label{eq:lbcstar}
    \inf_{d\in (0,1)} \frac{C(d)}{1-d}\geq \sup_{\lambda>0}\frac{C_\lambda}{\lambda}.
\end{equation}
The desired lower bound in Theorem~\ref{thm:poissonlb} follows by numerically lower bounding the right-hand side of~\eqref{eq:lbcstar} with recourse to the jigsaw decoding procedure from~\cite{DM07} applied to the Poisson-repeat channel with several different values of $\lambda$.

Given Theorem~\ref{thm:poissonlb}, it is natural to wonder what is the correct scaling of $C(d)$ with respect to $1-d$ when $d$ is close to $1$. In other words, we are interested in the quantity
\begin{equation}\label{eq:consthighdel}
    c^\star = \lim_{d\to 1^-} \frac{C(d)}{1-d}.
\end{equation}
For a complete proof of the existence of the limit on the right hand side of~\eqref{eq:consthighdel}, see~\cite{Dal10}. Theorem~\ref{thm:poissonlb} implies that
\begin{equation}\label{eq:boundscstar1}
    0.1185\leq c^\star\leq 1.
\end{equation}

The lower bound in~\eqref{eq:boundscstar1} is still the state-of-the-art. However, a significant effort was made to improve on the trivial upper bound. The first improvement on the upper bound was achieved by Diggavi, Mitzenmacher, and Pfister~\cite{DMP07}, who showed that
\begin{equation*}
    c^\star\leq 0.7918.
\end{equation*}

This result is obtained by coupling a marker-based approach analogous to the random process $V$ from~\cite{FD10} already discussed in Section~\ref{sec:lb} with numerical optimization methods. 
For deletion probability $d$, consider a modified deletion channel with markers inserted every $\ell=\frac{\lambda}{1-d}$ bits which are never deleted, where $\lambda$ is some constant (recall Figure~\ref{fig:regularmarkers} for an example). 
The markers can only increase the capacity, and this allows one to focus solely on the deletion channel with fixed input length $\ell$. 
Observe that the output length $L$ of this new channel follows a binomial distribution $\mathsf{Bin}(\ell,1-d)$ with mean $\lambda$. In the limit $d\to 1$, this means $L$ follows a $\Poi_{\lambda}$ distribution. 
The goal now is to numerically upper bound the capacity of this channel. Then, any numerical upper bound $u$ one obtains immediately yields the capacity upper bound for the standard deletion channel,
\begin{equation*}
    C(d)\leq \frac{u}{\ell}=\frac{(1-d)u}{\lambda}.
\end{equation*}
Hence, we conclude that $c^\star \leq u/\lambda$.

To numerically upper bound the capacity of the new channel, one wishes to apply Lemma~\ref{lem:AG}. 
In order to do this, however, additional modifications must be applied to the channel. First, the output of the channel must be truncated to strings of length at most $k$, where $k$ is some free constant. 
This truncation incurs an additive penalty in the capacity upper bound, stemming from the fact that the output length $L$ may be larger than $k$, at which point one may trivially upper bound the mutual information. More precisely, if $X$ is the channel input and $Y$ is the corresponding channel output, then
\begin{align*}
    I(X;Y) &= I(X;Y|L)\\
    &=\sum_{\ell=1}^\infty \Pr[L=\ell]\cdot I(X;Y|L=\ell)\\
    &\leq \sum_{\ell=1}^k \Pr[L=\ell]\cdot  I(X;Y|L=\ell)+\sum_{\ell=k+1}^\infty \Pr[L=\ell]\cdot \ell.
\end{align*}
Using the fact that $L$ is $\Poi_\lambda$, one can easily control the $\sum_{\ell=k+1}^\infty \Pr[L=\ell]\cdot \ell$ term and make it a small enough constant with appropriate choices of $\lambda$ and $k$. As desired, this allows one to consider only the channel with output length at most $k$ with a small penalty. Second, given this bound of $k$ on the output length, it is possible to show that it suffices to consider input $\ell$-bit sequences with at most $k$ runs to achieve capacity. This means it is possible to represent the input string as a vector $(q_1,\dots,q_k)$ with $\sum_{i=1}^k q_i=1$ and $q_i\geq 0$, where $q_i$ represents the relative length of the $i$-th run. These two observations combined with a careful implementation of the optimization algorithm with $\alpha=2$ and $k=6$ leads to the desired bound via Lemma~\ref{lem:AG}.

Later, using a different genie-aided argument, Fertonani and Duman~\cite{FD10} were able to improve the upper bound on $c^\star$ to
\begin{equation*}
    c^\star \leq 0.49.
\end{equation*}
To prove this bound, they consider the random process $W$ defined in Section~\ref{sec:ub} as side information to both sender and receiver in order to obtain a capacity upper bound.
It is then possible to take the limit of the resulting bound (normalized by $1-d$) as $d\to 1$ to derive the desired upper bound on $c^\star$.

Not long afterwards, Dalai~\cite{Dal10} presented yet another improved upper bound on $c^\star$, which has not been improved upon since.
\begin{thm}[\cite{Dal10}]\label{thm:ubcstar}
    We have $0.1185\leq c^\star\leq 0.4143$.
\end{thm}
Interestingly, the upper bound in Theorem~\ref{thm:ubcstar} is not obtained via a direct genie-aided argument. Instead, Dalai shows how to directly transform the current best upper bounds on $C(d)$ for \emph{general} $d$ into an upper bound on $c^\star$. More precisely, the main result in~\cite{Dal10} is the following.
\begin{thm}[\cite{Dal10}]\label{thm:infcstar}
    We have
    \begin{equation*}
        c^\star =\inf_{d\in(0,1)}\frac{C(d)}{1-d}.
    \end{equation*}
\end{thm}
Theorem~\ref{thm:ubcstar} is obtained from Theorem~\ref{thm:infcstar} by considering the best numerical capacity upper bound for the deletion channel~\cite{FD10} at $d=0.65$. Observe that any new improvements on upper bounds for $C(d)$ may directly yield an improved upper bound on $c^\star$. 
In order to prove Theorem~\ref{thm:infcstar}, Dalai uses the intuitive (and already mentioned) fact that, for large input length, the deletion channel with deletion probability $d$ is well-approximated by the exact deletion channel that deletes exactly $\lceil dn\rceil$ bits. 
Recall that the exact deletion channel already played a key role in the approach of Fertonani and Duman~\cite{FD10}. 
Then, Dalai exploits a known relationship between the exact deletion channel and $c^\star$ from~\cite{FD10}. 
Namely, recalling from Section~\ref{sec:ub} that $C_{\ell,r}$ denotes the capacity of the \emph{exact} deletion channel that receives $\ell$ bits as input and deletes exactly $\ell-r$ bits at random, for every $\ell\geq r$ it holds that
\begin{equation*}
    c^\star \leq \frac{\ell\cdot C_{\ell,r}+1}{r+1}.
\end{equation*}

We remark that Theorem~\ref{thm:ubcstar} can be also obtained as a corollary of the more general ``convexification'' result for the deletion channel capacity curve proved by Rahmati and Duman~\cite{RD15} which we discussed in Section~\ref{sec:ub}. For the sake of exposition and completeness, we chose to present Dalai's original approach to the $d\to 1$ setting here. In fact, recall that Rahmati and Duman proved that
\begin{equation}\label{eq:convexification}
    C(\lambda d+(1-\lambda))\leq \lambda C(d)
\end{equation}
for all $d,\lambda\in[0,1]$. 
For all fixed $d^\star\in [0,1]$ and $d\geq d^\star$ (setting $\lambda=\frac{1-d}{1-d^\star}$), this readily implies that
\begin{equation}\label{eq:ubconvexification}
    \frac{C(d)}{1-d}\leq \frac{C(d^\star)}{1-d^\star}.
\end{equation}
Analogously to the previous approach, setting $d^\star=0.65$ and using the best known upper bound for that value leads to Theorem~\ref{thm:ubcstar}. Note, however, that this is a stronger result than the general upper bound obtained via Theorem~\ref{thm:infcstar} as it holds for all $d\geq d^\star$, not only in the limit $d\to 1$.

A common feature of all previous bounds on $c^\star$ is that they crucially depend on heavy numerical computations. However, by combining~\eqref{eq:ubconvexification} with results derived in~\cite{Che19}, it is possible to obtain upper bounds on $c^\star$ that can be derived without computer assistance. For example, starting with the fully explicit upper bound
\begin{equation*}
    C(1/2)\leq \frac{\log \phi}{2}
\end{equation*}
from~\cite{Che19}, where $\phi\approx 1.618$ denotes the golden ratio, we conclude that
\begin{equation*}
    c^\star \leq 0.6942
\end{equation*}
without computer assistance, improving on the first non-trivial bound on $c^\star$ from~\cite{DMP07}.

Collecting the state-of-the-art results on $c^\star$, we have
\begin{equation*}
    0.1185\leq c^\star\leq 0.4143.
\end{equation*}
This suggests some natural open problems:
\begin{itemize}
    \item Narrow the gap between lower and upper bounds for $c^\star$;
    
    \item Both state-of-the-art bounds require computer assistance to be determined. With this in mind, it would be interesting to get close (or even surpass) both bounds with a completely explicit approach;
    
    \item All the results presented guarantee only with existential results for codebooks. It would be interesting to construct efficiently encodable and decodable codes with rates close to the best known lower bound on $c^\star$ for large deletion probability.
\end{itemize}

\paragraph{Is the deletion channel capacity curve convex?}

A natural and seemingly more approachable conjecture put forth by Dalai~\cite{Dal10} about the deletion channel capacity is that this curve is convex in the interval $[0,1]$. Such a result would not only lead to a new qualitative understanding of the deletion channel, but it would also allow us to immediately extend numerical upper bounds beyond specific, isolated values of $d$ to all $d\in[0,1]$ (with small loss in the quality of the bound for $d$ close to the points at which the original bound was computed). Some known results support this conjecture. Namely, Theorem~\ref{thm:infcstar} by Dalai and the convexification result obtained by Rahmati and Duman are necessary (but not sufficient) conditions for $C(d)$ to be convex.
We remark also that the capacity bounds we currently have for the deletion channel are also convex (recall Figure~\ref{fig:capdel}).

Interestingly, this conjecture turns out to be false in the finite blocklength regime. Let $C_n(d)$ denote the capacity of the deletion channel with fixed input length $n$. As observed by Dalai~\cite{Dal10}, as $n$ increases it can be experimentally observed that $C_n(d)$ is convex in a neighborhood of $d=0$ of increasing size, but it is always concave in a neighborhood of $d=1$.



\section{Synchronization channels concatenated with memoryless channels }\label{sec:noisy}

\begin{figure}
	\centering
	\includegraphics[width=0.5\textwidth]{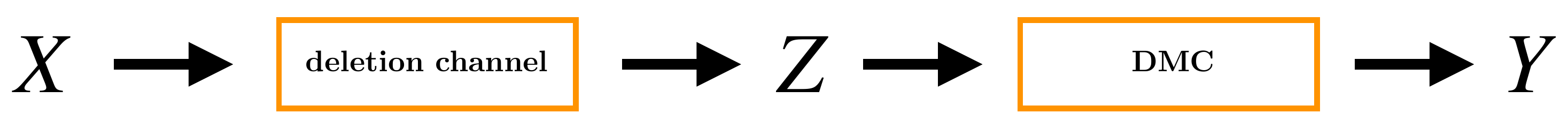}
	\caption{The channel obtained by concatenating a deletion channel with a DMC.}
	\label{fig:concat}
\end{figure}

Several works have considered communication under the joint effect of synchronization loss and memoryless errors, such as substitutions (``bit-flips''), erasures, and Additive White Gaussian Noise (AWGN). Here, we focus our exposition mainly on the concatenation of a deletion channel with a DMC.
Figure~\ref{fig:concat} illustrates the resulting channel.

We remark that the works discussed below also present results for more general synchronization channels. Moreover, although the channels considered here do not fall directly into the class considered in Section~\ref{sec:shannontype}, a similar argument guarantees a result analogous to Theorem~\ref{thm:capsynch}. 
Namely, their capacity is also given by
\begin{equation*}
    \lim_{n\to\infty}\max_{X^{(n)}}\frac{I(X^{(n)};Y_{X^{(n)}})}{n},
\end{equation*}
where the maximum is taken over all distributions $X^{(n)}$ supported on $\cX^n$.

This setting was first studied by Gallager~\cite{Gal61}, who introduced a channel (over a binary alphabet) where bits are first deleted with probability $p_d$ or replaced by \emph{two} random bits with probability $p_i$. Bits which are neither deleted nor replaced by two random bits (which happens with probability $1-p_d-p_i$) are then flipped with probability $p_e$. Using convolutional coding techniques, Gallager~\cite{Gal61} showed that the capacity of this channel is lower bounded by
\begin{equation}\label{eq:galLB}
    1+p_d\log p_d+p_i\log p_i+p_e\log p_e+p_c\log p_c+p_f\log p_f
\end{equation}
with the understanding that $0\log 0=0$, where $p_c=(1-p_d-p_i)(1-p_e)$ is the probability of correct transmission and $p_f=(1-p_d-p_i)p_e$ is the probability a bit survives the first stage but is flipped in the second stage. Observe that when $p_i=0$ this channel corresponds to the concatenation of a deletion channel with deletion probability $p_d$ and a BSC with error probability $p_e$, which we call the \emph{deletion-BSC channel}.
Denoting the capacity of this channel by $C(p_d,p_e)$, from~\eqref{eq:galLB} we obtain the capacity lower bound
\begin{equation}\label{eq:lbDelBSC}
    C(p_d,p_e)\geq 1-h(p_d)-(1-p_d)\cdot h(p_e).
\end{equation}

A more general approach was later undertaken by Diggavi and Grossglauser~\cite{DG06}, who considered the concatenation of a deletion channel and an arbitrary DMC. Using a uniform input distribution combined with a typicality argument similar to that used to derive~\eqref{eq:lbBSC},
they show that the capacity of the concatenation of a deletion channel with probability $d$ and a DMC $\Ch$, which we denote by $C(d,\Ch)$, satisfies
\begin{equation*}
    C(d,\Ch)\geq 1-h(d)-(1-d)H(Z|Y),
\end{equation*}
where $Z$ is a uniform input to the DMC and $Y$ is the output distribution of the DMC on input $Z$. If the DMC is \emph{symmetric} (in particular, the capacity of such DMC's is achieved by a uniform input distribution), then we have
\begin{equation*}
    C(d,\Ch)\geq 1-h(d)-(1-d)(1-\Ca(\Ch)).
\end{equation*}
In particular, if $\Ch$ is a BSC, we recover~\eqref{eq:lbDelBSC}.
As another example, if $\Ch$ is a BEC with erasure probability $\eps$, we conclude that
\begin{equation}\label{eq:lbdelbec}
    C(d,\Ch)\geq 1-h(d)-(1-d)\eps.
\end{equation}
On the other hand, one also has the straightforward upper bound
\begin{equation*}
    C(d,\Ch)\leq (1-d)\Ca(\Ch)
\end{equation*}
obtained by considering a genie that reveals the positions of all deleted bits to the receiver.

There is a simple, general observation exploited by Rahmati and Duman~\cite{RD14} that opens the door to improved lower bounds. Suppose $X$ is the $n$-bit input distribution to a concatenation of a deletion channel\footnote{The argument that follows works with any synchronization channel, but in this section we pay particular attention to the deletion channel.} with a DMC $\Ch$.
Let $Z$ be the output of the deletion channel on input $X$, and $Y$ the output of $\Ch$ on input $Z$ (as in Figure~\ref{fig:concat}). Then, if we have the inequality
\begin{equation*}\label{eq:DMCdelmutinf}
    I(X;Y)\geq I(X;Z)-an
\end{equation*}
for every such tuple $X\rightarrow Z\rightarrow Y$ and a fixed constant $a$, it easily follows that
\begin{equation*}\label{eq:DMCdel}
    \Ca(d,\Ch)\geq C(d)-a.
\end{equation*}
This approach allows one to directly leverage the best lower bounds for the deletion channel to obtain good lower bounds for its concatenation with DMC. For example, instantiating $\Ch$ as a BSC with error probability $\eps$, it holds that~\cite{RD14}
\begin{equation*}
    I(X;Y)\geq I(X;Z)-n(1-d)\cdot h(\eps)
\end{equation*}
for every $n$-bit input distribution $X$. As per the discussion above, this means that
\begin{equation*}
    C(d,\eps)\geq C(d)-(1-d)\cdot h(\eps).
\end{equation*}
This inequality immediately leads to improved capacity lower bounds when compared to~\eqref{eq:lbDelBSC}, both numerical (by using the state-of-the-art lower bounds from~\cite{RD15}), or analytical (since there are better analytical lower bounds on $C(d)$ than $1-h(d)$, see~\cite{DG06,RD13}). Rahmati and Duman~\cite{RD14} successfully apply the observation above to obtain lower bounds beyond what we present here. In fact, they reduce the task of deriving capacity lower bounds for \emph{general} synchronization channels concatenated with DMC's combining erasures and bit-flips and AWGN to that of obtaining good lower bounds for the underlying synchronization channel.

Although we have been focusing solely on capacity lower bounds thus far, capacity upper bounds also exist for some channels of this type. Fertonani, Duman, and Erden~\cite{FDE11}, based on techniques from~\cite{FD10} already discussed in Section~\ref{sec:lb}, derived numerical capacity upper and lower bounds for concatenations of a deletion/insertion channel and a BSC.
Recall the random process $V$ discussed in Section~\ref{sec:lb}.
Then, for any fixed integer $\ell$, one can follow an analogous reasoning to the one behind the derivation of~\eqref{eq:cellbounds} to obtain the bounds
\begin{equation*}
    C'_\ell \geq C(d,\eps)\geq C'_\ell-\frac{1}{\ell}H(V_1).
\end{equation*}
The only difference with respect to the derivation of~\eqref{eq:cellbounds} in Section~\ref{sec:lb} is that instead we have
\begin{equation*}
    C'_\ell = \max_{X^{[1]}} \frac{1}{\ell} I(X^{[1]};Y^{[1]})
\end{equation*}
for $X^{[1]}$ supported over $\bits^\ell$ and $Y^{[1]}$ its corresponding output through the \emph{deletion-BSC channel} with deletion probability $d$ and bit-flip probability $\eps$ (instead of the deletion channel as before).
If $\ell$ is a small constant, then $C'_\ell$ can be directly approximated to good accuracy with computer assistance via the Blahut-Arimoto algorithm (in~\cite{FDE11}, the authors go up to $\ell=8$).
We remark that this approach yields good numerical capacity upper bounds for a concatenation of a deletion channel with many DMC's, and not just a BSC.

The capacity upper bound for the deletion-BSC channel derived in~\cite{FDE11} is not convex for deletion probability $d\geq 0.6$.
Rahmati and Duman~\cite{RD15} show that their ``convexification'' result for the standard deletion channel obtained via channel fragmentation (discussed in Section~\ref{sec:ub}) can be extended to the deletion-BSC channel.
In particular, they proved that
\begin{equation*}
    C(\lambda d^\star+1-\lambda,\eps)\leq \lambda C(d^\star,\eps)
\end{equation*}
holds for all $\lambda,d^\star\in[0,1]$ and $\eps\in[0,1/2]$.
Setting $\lambda=\frac{1-d}{1-d^\star}$ yields the capacity upper bound
\begin{equation}\label{eq:convexnoisy}
    C(d,\eps)\leq \frac{1-d}{1-d^\star}C(d^\star,\eps)
\end{equation}
for every $d\geq d^\star$.
Intuitively, the inequality in~\eqref{eq:convexnoisy} means that every capacity upper bound for fixed $d^\star$ can be linearly extended to all $d\geq d^\star$.
This allows Rahmati and Duman to improve upon the upper bounds from~\cite{FDE11} for $d>0.6$ by considering the upper bound obtained in~\cite{FDE11} at $d^\star=0.6$ and applying~\eqref{eq:convexnoisy} for all $d\geq 0.6$.

We remark that synchronization channels concatenated with DMC's have also been studied in the context of nanopore sequencing of DNA strands by Mao, Diggavi, and Kannan~\cite{MDK18}. This topic recently found close connections to portable DNA-based data storage~\cite{YGM17,OAC+18}. In~\cite{MDK18}, the authors consider a channel model combining inter-symbol interference, deletions, and memoryless errors. They prove an analogue of Theorem~\ref{thm:capsynch} for their setting, and, using techniques similar to those presented above, derive capacity bounds.


Finally, besides the capacity bounds discussed above, we note that some works have also derived simulation-based capacity bounds for several types of synchronization channels with memoryless errors~\cite{ZTMK05,HDEK10,MTL12}. Similarly to the case of synchronization channels without memoryless errors discussed in Sections~\ref{sec:lb} and~\ref{sec:ub}, these works consider Markov input sources and estimate their information rate.


\paragraph{The large alphabet setting}
Similarly to what has been discussed in Sections~\ref{sec:lb} and~\ref{sec:ub}, although \emph{binary} synchronization channels concatenated with DMC's are hard to handle, the problem of how the capacity scales with input alphabet size is more approachable. For example, Mercier, Tarokh, and Labeau~\cite{MTL12} showed that the capacity of the $Q$-ary deletion channel with deletion probability $d$ concatenated with a $Q$-SC channel\footnote{A $Q$-SC channel is a generalization of the BSC to a $Q$-ary alphabet. For every $q,q'\in[Q]$ such that $q\neq q'$, an input symbol $q$ is corrupted to $q'$ with probability $\frac{\eps}{Q-1}$, and is transmitted correctly with probability $1-\eps$.} with error probability $\eps$, which we denote by $C_Q(d,\eps)$, behaves like
\begin{equation*}
    C_Q(d,\eps)\sim (1-d)(1-\eps)\log Q
\end{equation*}
as $Q$ increases.
Moreover, this optimal scaling is achieved by a uniform input distribution.
They also determine the exact scaling of the capacity with respect to the alphabet size for other (more complex) types of synchronization channels combining deletions, geometric insertions and duplications, and memoryless errors.

\section{Multi-use synchronization channels}\label{sec:multiuse}

Although we have been studying settings where an input string is sent through a synchronization channel only once, it is of both practical and theoretical interest to analyze the case where one is allowed to send a string several times through a channel. 

The multi-use setting for synchronization channels, in particular for channels with deletions and insertions, is practically motivated by the read process in portable DNA-based data storage systems~\cite{YGM17,OAC+18}. In DNA-based storage, data is encoded into DNA strands. To read the data off a given strand, portable systems make use of so-called \emph{nanopore sequencers}. Roughly speaking, the DNA strand in question is sent through several pores, and each pore produces a noisy read of its content, inducing deletions, insertions, and substitutions. Therefore, it is important to 
understand the fundamental limits of reliable communication with a given number of noisy reads.

The multi-use deletion channel is also closely related to the problem of trace reconstruction, first studied by Levenshtein~\cite{Lev01,Lev01b} from a combinatorial perspective and Batu, Kannan, Khanna, and McGregor~\cite{BKKM04} for i.i.d.\ errors, motivated by applications in computational biology, phylogeny, and sequence alignment. The differences between the two problems are that in trace reconstruction the number of channel uses is not constant (it may grow with the blocklength), and the input string is either arbitrary or uniformly random. We note, however, that the problem of coding for the multi-use deletion channel with non-constant number of uses has been recently considered~\cite{AVDF19,CGMR19,BLS19}, again with connections to DNA-based data storage.

From a purely information-theoretic viewpoint, it is also interesting to compare how the capacity of synchronization channels behaves with respect to the number of uses versus the same scaling for DMC's. The multi-use capacity of many DMC's can be easily bounded, while the same cannot be said for similar channels with memory. For example, consider the multi-use BEC. Using the simple observation that a BEC with erasure probability $d$ and $t$ uses is equivalent to a BEC with erasure probability $d^t$, it follows immediately that its $t$-use capacity equals $1-d^t$. However, the situation for its close neighbor, the deletion channel, appears to be much more complex, since we do not know of any simple connection between the single-use and multi-use versions. 

Haeupler and Mitzenmacher~\cite{HM14} studied the multi-use deletion channel in the regime of small deletion probability. Namely, they were able to determine the first-order term of the capacity in this regime \emph{under a uniform codebook}, i.e., when all codewords are chosen uniformly at random. Interestingly, this quantity is close to the capacity of the $t$-use BSC, extending the close connection between the single-use deletion channel and the BSC in the small deletion probability regime discussed in Section~\ref{sec:lowdel} to multiple uses.
\begin{thm}[\cite{HM14}]\label{thm:multiuse}
    For any integer $t>0$, the capacity $C(t,d)$ of the $t$-use deletion channel with deletion probability $d$ under a random codebook satisfies
    \begin{equation*}
        C(t,d)= 1 - A(t)\cdot d^t\log(1/d)-O(d^t),
    \end{equation*}
    where
    \begin{equation*}
        A(t)=\sum_{j=1}^{\infty}2^{-j-1}tj^t.
    \end{equation*}
\end{thm}

As described in~\cite{HM14}, the idea behind the proof of Theorem~\ref{thm:multiuse} comes from a simple strategy towards bounding the capacity of the multi-use BSC. Note that we cannot immediately reduce the multi-use BSC to the single-use case as we did for the BEC. However, such a reduction works if we first resolve some of the errors with a genie-based argument. First, observe that we can always tell whether some input bit was flipped in some of the $t$ BSC outputs provided that it was not flipped in every use of the channel. If this exception happens, then we cannot detect that any bit flips occurred, and hence one calls this an \emph{undetectable error}. Clearly, if the bit-flip probability is $d$, an undetectable error occurs at a given position with probability $d^t$. If these were the only errors we had to worry about, then the resulting channel would be equivalent to a single-use BSC with bit-flip probability $d^t$, whose capacity we know. Although this is not exactly the case, one can show that the capacity of the true multi-use BSC is close to the capacity of the channel with only undetectable errors whenever the bit-flip probability $d$ is small enough. To see this, consider the $t$-use BSC equipped with a genie that, for each detectable error that occurs, tells the receiver if at least $t/2$ bit-flips occurred. The information from the genie allows the receiver to correct every detectable error, leading to the desired channel with only undetectable errors. To show the capacities are close for small $d$, it is now enough to argue that, on average, the genie must reveal very few bits of side information. If this holds, then the addition of the genie affects only lower-order terms of the capacity. In the BSC case this is easy to show, since a standard concentration argument shows the probability that more than $t/2$ bit-flips occur (at which point the genie must tell the receiver) in a given position is small enough with respect to $d$ and $t$.

\begin{figure}
	\centering
	\includegraphics[width=0.5\textwidth]{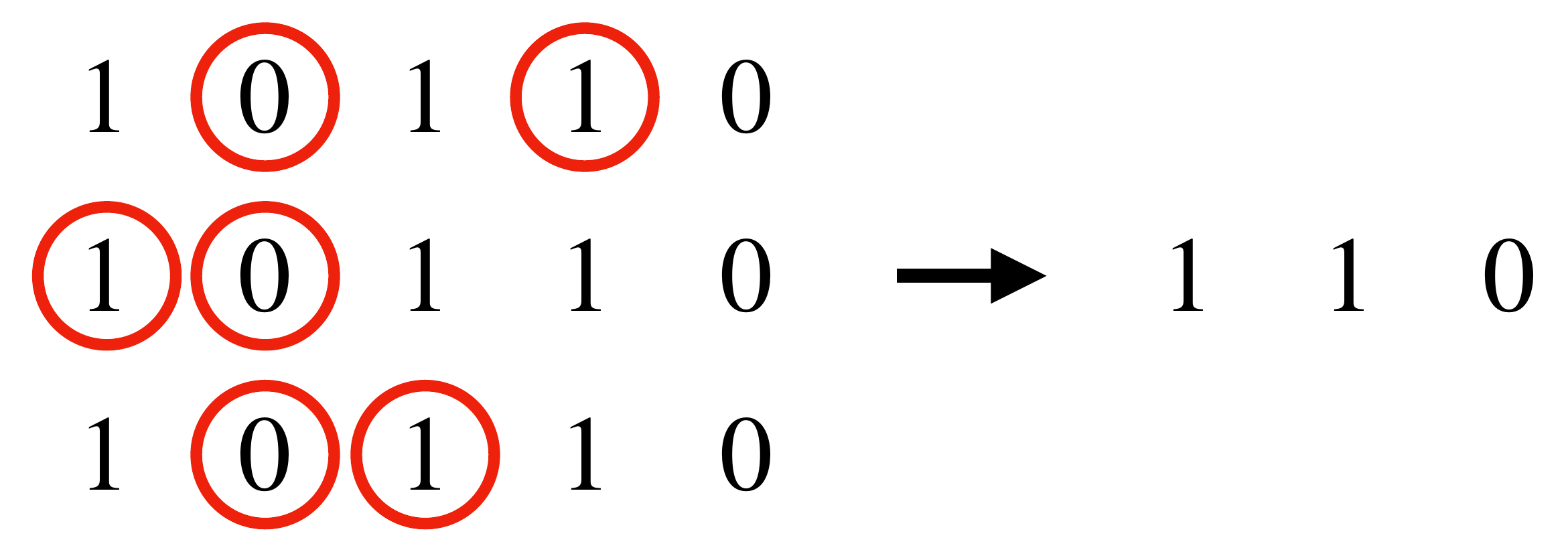}
	\caption{A set of undetectable deletions. Circled bits are deleted. Note that every set of errors that leads to the same output is undetectable.}
	\label{fig:undetectable}
\end{figure}

The idea used by Haeupler and Mitzenmacher~\cite{HM14} for the deletion channel is similar to that described in the paragraph above. However, more care is needed. First, the set of undetectable errors is not as simple as for the BSC. Analogously to the BSC, the situation where the $i$-th bit is deleted in all $t$ uses leads to an undetectable error. However, other deletion patterns may also lead to undetectable errors (see Figure~\ref{fig:undetectable} for an example). At a high level, Haeupler and Mitzenmacher~\cite{HM14} proceed by characterizing the set of undetectable errors, showing that they occur with probability $O(d^t)$, and finally showing that, on average, the genie must reveal only $O(d^t)$ bits of side information per input bit to resolve all other errors. This allows one to effectively approximate the $t$-use deletion channel by the channel that inflicts only undetectable errors when the deletion probability $d$ is small enough. Determining the capacity of the latter turns out to be approachable.

We note that the multi-use setting for sticky channels has been considered by Magner, Duda, Szpankowski, and Grama~\cite{MDSG16}. This is motivated by nanopore sequencing, which, as mentioned before, has recently found connections to DNA-based data storage~\cite{YGM17,OAC+18}. However, they do not study the capacity directly, and their focus lies on the case with non-constant number of channel uses (i.e., the trace reconstruction problem~\cite{BKKM04}). Notably, they show that for well-behaved replication distributions $\log n$ channel uses suffice for reliable reconstruction of typical sequences, where $n$ is the input block length.

We conclude this section with a list of open problems:
\begin{itemize}
    \item Obtain non-trivial capacity bounds for the $t$-use deletion channel beyond the small deletion probability regime;
    
    \item Consider channels that introduce more types of errors (e.g., insertions and deletions) on top of deletions;
    
    \item Design good codes for the multi-use deletion channel.
\end{itemize}

\section{Zero-rate threshold of adversarial synchronization channels}\label{sec:zerorate}

In contrast with previous sections, the current section deals with reliable information transmission under \emph{adversarial} synchronization errors. The main focus will be on adversarial binary deletions, although the effect of insertions and replications over a binary alphabet will also be considered.

In an adversarial setting, the most relevant notion is that of \emph{zero-error capacity}. More precisely, we consider a setting where an adversary is allowed to corrupt a codeword of length $n$ with at most $dn$ synchronization errors (with $0\leq d <1$), and we are interested in the size of the largest codebook that allows for decoding with zero probability of error under this model. More precisely, one wishes to determine the supremum of all rates $R\geq 0$ such that for large enough $n$ there is a codebook $\cC\subseteq \mathcal{X}^n$ of size $|\cC|=|\mathcal{X}|^{Rn}$ coupled with a decoder $\Dec:\cY^*\to\mathcal{X}^n$ satisfying the property that, if $c\in\cC$ and $c'\in\cY^*$ can be obtained by corrupting $c$ with at most $pn$ synchronization errors, then $\Dec(c')=c$.

Determining the zero-error capacity of the adversarial binary deletion channel with respect to the fraction of deletions $d$, which we denote by $C_{\mathsf{adv, del}}(d)$, for non-trivial $d$ is a very challenging problem. Therefore, previous works have focused mostly on the easier problem of understanding zero-rate \emph{threshold} for the adversarial deletion channel. More precisely, we seek to determine the fraction $d^\star$ such that $C_{\mathsf{adv, del}}(d)>0$ for all $d<d^\star$, and $C_{\mathsf{adv, del}}(d)=0$ for all $d>d^\star$.

We begin our discussion with a simple remark, which is an immediate consequence of the fact that, if we allow at least $n/2$ adversarial deletions, then an adversary can force the channel to output either all $0$'s or all $1$'s.
\begin{remark}
The zero-rate threshold of the adversarial deletion channel, $d^\star$, satisfies $d^\star\leq 1/2$.
\end{remark}

It is also not clear at first whether $d^\star>0$. This was established by Schulman and Zuckerman~\cite{SZ99}, who constructed a code with positive rate handling some constant fraction of adversarial deletions and insertions (besides another type of errors not relevant for our discussion). Notably, their code not only achieves positive rate, but also supports efficient (i.e., running in time polynomial in the blocklength of the code) encoding and decoding. Their construction is based on code concatenation. The outer code is a slightly modified Reed-Solomon code where codeword symbols consist in evaluation points concatenated with their evaluation under the message polynomial. The binary inner code has logarithmic block length, and hence can be constructed in a greedy manner to satisfy the desired properties (codewords should be sufficiently far apart in edit distance, and should have at least half $1$'s in every block of even length) efficiently in the final blocklength.

Improved bounds on $d^\star$ were derived by Kash, Mitzenmacher, Thaler, and Ullman~\cite{KMTU11} by relating $d^\star$ with the expected length of the longest common subsequence (LCS) of two independent, uniformly random $n$-bit strings, denoted by $L_n$. Observe that a code cannot be zero-error decoded from $dn$ adversarial deletions if and only if there exist two codewords whose longest common subsequence has length at least $(1-d)n$. This suggests the following approach towards showing the existence of a code that can be zero-error decoded from at most $dn$ adversarial deletions: Consider the graph $G_{n,d}$ with vertices in $\{0,1\}^n$ and an edge between $x$ and $y$ whenever the length of the LCS between $x$ and $y$, denoted by $\LCS(x,y)$, satisfies $\LCS(x,y)\geq (1-d)n$. Then, every independent set of $G_{n,d}$ yields a code with the desired property. If $G_{n,d}$ has a large enough independent set for all $n$ large enough, then we obtain the lower bound $d\leq d^\star$.

Suppose that $1-d>\gamma$, where
\begin{equation}\label{eq:limlcs}
    \gamma = \lim_{n\to\infty}\frac{L_n}{n}.
\end{equation}
Intuitively, $\gamma$ is the relative expected length of the longest common subsequence of two uniformly random $n$-bit strings for large $n$. Recall that our goal is to show the existence of a large enough independent set in $G_{n,d}$. 
One way to do this is to first show that $G_{n,d}$ has few edges. 
Then, we can apply Tur\'an's theorem~\cite{Tur41} to show existence of the desired independent set\footnote{Tur\'an's theorem states that any graph with $n$ vertices and $e$ edges contains an independent set of size at least $\frac{n^2}{2e+n}$.}.
This idea is easy to realize: First, for large $n$ and uniformly random strings $X$ and $Y$, it holds that $\LCS(X,Y)$ is well-concentrated around $\gamma n$. 
More precisely, a standard concentration argument shows that all but an exponentially small fraction of pairs of strings $x$ and $y$ satisfy
\begin{equation*}
    |\LCS(x,y)-\gamma n|<(1-d)n-\gamma n.
\end{equation*}
As a result, since there is an edge between $x$ and $y$ only when $\LCS(x,y)\geq (1-d)n$, the number of edges in $G_{n,d}$ is at most $\binom{n}{2}2^{-cn}$ for some constant $c>0$ depending only on $d$. Tur\'an's theorem then guarantees the existence of an independent set in $G_{n,d}$, and hence a code, of size $2^{c'n}$ for some constant $c'>0$. Since this code has rate at least $c'>0$, we have the following result.
\begin{thm}[\cite{KMTU11}]\label{thm:thresholdlcs}
    It holds that $d^\star\geq 1-\gamma$.
\end{thm}

\begin{remark}
The limit on the right hand side of~\eqref{eq:limlcs} can be shown to exist via subadditivity and Fekete's lemma (see Lemma~\ref{lem:fekete}).
\end{remark}

Deriving good upper bounds on $\gamma$ is a challenging problem in combinatorics. The best result, due to Lueker~\cite{Lue09}, shows that $\gamma\leq 0.8263$. This leads to the following corollary of Theorem~\ref{thm:thresholdlcs}.
\begin{coro}[\cite{KMTU11}]\label{coro:thresholdlcs}
    It holds that $d^\star\geq 0.1737$.
\end{coro}

We briefly discuss the limitations of the approach undertaken in~\cite{KMTU11}. The best lower bound known for $\gamma$ is $0.788$~\cite{Lue09}. This means that the best lower bound on $d^\star$ one may hope to prove with the approach above is $d^\star\geq 0.222$. Nevertheless, as considered in~\cite{KMTU11}, it is possible to generalize the previous connection and relate $d^\star$ to the expected length of the LCS of two strings generated by independent copies of any first-order Markov chain. Although no rigorous bounds are known for this quantity, numerical simulations suggest that one should be able to derive a lower bound close to $0.25$ for $d^\star$ with this generalized approach.

It is instructive to compare the observations in the paragraph above with known results for channels with adversarial bit flips. In this error model, it is known that the zero-error capacity against $dn$ bit-flips is zero for $d>0.25$. 
This leads to the following natural question: 
Is the zero-rate threshold for the adversarial deletion channel also $d^\star=0.25$? This turns out not to be the case. 
A significant improvement on the lower bound from Corollary~\ref{coro:thresholdlcs} was obtained by Bukh, Guruswami, and H{\aa}stad~\cite{BGH17} using different techniques. 
Interestingly, they showed that the adversarial deletion channel is \emph{significantly easier} to handle than the adversarial binary symmetric channel. More precisely, they proved the following.
\begin{thm}[\cite{BGH17}]\label{thm:thresholdlb}
    It holds that $d^\star\geq \sqrt{2}-1\approx 0.4142$.
\end{thm}

We proceed to give a high-level description of a construction from~\cite{BGH17} that achieves the lower bound in Theorem~\ref{thm:thresholdlb}. Although the code we present is not efficiently encodable nor decodable, we remark that the lower bound can in fact be achieved with a family of efficiently encodable and decodable codes~\cite{BGH17}.

Recall that, for any $d\in (0,1)$, we are aiming to show the existence of a binary code $\cC$ with positive rate which ensures that $\LCS(c,c')<(1-d)n$ for all distinct codewords $c,c'\in\cC$. This leads to the lower bound $d\leq d^\star$. The starting point of the construction from~\cite{BGH17} is the observation that if we allow the alphabet size $Q$ to be large enough, then a standard probabilistic argument shows that a uniformly random code $\cC\subseteq [Q]^n$ with positive rate satisfies $\LCS(c,c')\leq \gamma n$ for any $c,c'\in\cC$ and $\gamma$ much smaller than $1-d$. We now want to transform this large-alphabet code with strong guarantees into a binary code with a good enough constraint on the LCS between any two distinct codewords. 
This is achieved by concatenating $\cC$ with an appropriate binary inner code.

The starting point towards defining the correct inner code is a family of codes first introduced and studied by Bukh and Ma~\cite{BM14}. The Bukh-Ma codes~\cite{BM14} correspond to sets of codewords of the form
\begin{equation}\label{eq:bukhma}
    \{(0^r 1^r)^{\frac{n}{2r}}:r\in\mathcal{R}\},
\end{equation}
where $\mathcal{R}\subseteq\mathbb{N}$ is a finite set of integers sufficiently far apart. At a high level, this is required to ensure that the LCS between any two distinct codewords is not very large.
Combining a Bukh-Ma inner code with the random outer code already leads to the non-trivial lower bound $d^\star\geq 1/3$, improving upon Theorem~\ref{thm:thresholdlcs}.
However, to improve the lower bound on $d^\star$, one needs to consider a modified version of the Bukh-Ma codes where long runs of $0$'s are periodically interspersed with short runs of $1$'s and vice-versa. More specifically, for a constant $0\leq c<1$, let
\begin{equation*}
    0_{r,a}=(0^a 1^{ca})^{\frac{r}{(1+c)a}}
\end{equation*}
and
\begin{equation*}
    1_{r,a}=(0^{ca} 1^{a})^{\frac{r}{(1+c)a}}.
\end{equation*}
Then, the modified Bukh-Ma code is defined as
\begin{equation*}
    \{(0_{r,a} 1_{r,a})^{\frac{n}{2r}}:r\in\mathcal{R}, a\in\cA\},
\end{equation*}
where both $\mathcal{R}$ and $\cA$ are finite sets of integers appropriately far apart. Note that the original Bukh-Ma codes are a special case of these codes obtained by setting $c=0$.
Adding periodic short runs (of sufficiently different lengths between any two distinct codewords) allows to decrease the length of the LCS between any two distinct codewords, improving the lower bound on $d^\star$ from $1/3$ to $\sqrt{2}-1$.

\paragraph{The zero-rate threshold for larger alphabets}

Although we have focused on the binary setting only, we note that some works also studied the zero-rate threshold of the adversarial deletion channel over larger alphabets. For an alphabet of size $Q$, denote its zero-rate threshold by $d^\star_Q$. Then, similarly to the binary setting, the upper bound $d^\star_Q\leq 1-1/Q$ is immediate -- simply delete all occurrences of the most common symbol in the codeword. The constructions of Bukh, Guruswami, and H{\aa}stad~\cite{BGH17} can be generalized to larger alphabets to show that $d^\star_Q\geq 1-\frac{2}{Q+\sqrt{Q}}$ for all $Q\geq 2$. In particular, this shows that the zero-rate threshold asymptotically scales like $1-\Theta(1/Q)$. This improves on the previous best lower bound for large alphabets of $d^\star_Q\geq 1-O(1/\sqrt{Q})$. As remarked by Guruswami and Wang~\cite{GW17}, this worse bound can be obtained by combining the approach used in~\cite{KMTU11} with the fact that the length of the LCS between two random strings in $[Q]^n$ scales like $2n/\sqrt{Q}$ proved in~\cite{KLM05}.

\paragraph{The zero-rate threshold of list-decoding from deletions and insertions}

In the remainder of this section, we focus on the zero-rate threshold in relaxed settings. The first such relaxation consists in requiring only that the desired code $\cC\subseteq \bits^n$ be \emph{$L$-list-decodable} from a combination of synchronization errors (namely deletions and insertions), where $L=\textrm{poly}(n)$.\footnote{A code $\cC$ is said to be $L$-list-decodable from $pn$ synchronization errors if for every pair $(c,c')$ such that $c\in\cC$ and $c'$ is obtained from $c$ by introducing at most $pn$ errors, it holds that there are at most $L$ codewords $\hat{c}\in\cC$ that could lead to output $c'$ after the introduction of at most $pn$ errors.} This stands in contrast with the unique decodability requirement in our previous discussion. Another departure from previous sections consists in the fact that we will be considering both (adversarial) deletions and insertions.

We begin by considering the deletion-only list-decoding setting first. Denote the zero-rate threshold for $Q$-ary codes in this case by $d^\star_{Q,L}$. Then, our first simple observation is that the $1-1/Q$ upper bound from the unique decoding deletion-only setting also extends to list-decoding. In other words, we have $d^\star_{Q,L}\leq 1-1/Q$. This readily implies, given our previous results on the zero-rate threshold under unique decoding, that $d^\star_{Q,L}=1-\Theta(1/Q)$. However, this does not tell us much about threshold for small alphabets. In particular, when $Q=2$, can we prove stronger results about $d^\star_{2,L}$ than what we know about $d^\star$?

Before we consider this question, we briefly discuss the zero-rate threshold in the case of insertions only. Note that unique decoding from insertions only is equivalent to unique decoding from deletions. This is due to the fact that a code corrects $d$ adversarial deletions if and only if it corrects the same number of insertions~\cite{Lev65}. As a result, the zero-rate threshold is the same in both cases. 
However, this equivalence does not extend to the list-decoding setting. 
The first progress on the zero-rate threshold for list-decoding from insertions was made by Hayashi and Yasunaga~\cite{HY18}, who showed that the codes from~\cite{BGH17} already used to prove Theorem~\ref{thm:thresholdlb} are also $(L=\textrm{poly}(n))$-list-decodable from a $0.707$-fraction of adversarial insertions. 
Denoting the zero-rate threshold from insertions in the list-decoding setting (with binary alphabet) by $d^\star_{2,L,\mathsf{ins}}$, this implies that $d^\star_{2,L,\mathsf{ins}}\geq 0.707$. We note also that the upper bound $d^\star_{2,L,\mathsf{ins}}\leq 1$ is easy to prove. Indeed, it is enough to observe that $n$ insertions suffice to transform any $n$-bit string into the $2n$-bit string $0101\dots 01$. We are then left with question of where $d^\star_{2,L,\mathsf{ins}}$ lies between $0.707$ and $1$.

Recently, Guruswami, Haeupler, and Shahrasbi~\cite{GHS19} answered both questions above in full. Notably, besides these results, they settle the zero-rate threshold of list-decoding from a \emph{combination} of deletions and insertions for all alphabet sizes $Q$ (for polynomial list-size). We present their result for $Q=2$ only.
\begin{thm}[\cite{GHS19}]\label{thm:binlistdec}
	For any $\eps>0$ and all $n$, there exists an efficiently encodable and decodable code $\cC\subseteq \bits^n$ with positive rate that is $\textrm{poly}(n)$-list-decodable from a $\delta$-fraction of deletions and a $\gamma$-fraction of insertions whenever $2\delta+\gamma\leq 1-\eps$. In particular, we have $d^\star_{2,L}=1/2$ and $d^\star_{2,L,\mathsf{ins}}=1$.
\end{thm}
Interestingly, the analogous result for general alphabet size $Q$ is not an easy generalization of the statement above. 
It is straightforward to check that Theorem~\ref{thm:binlistdec} is tight, i.e., no such positive-rate codes exist when $2\delta+\gamma\geq 1$. 

Theorem~\ref{thm:binlistdec} is proved via code concatenation.
As the outer code, Guruswami, Haeupler, and Shahrasbi choose one of the high-rate codes from~\cite{HSS18} which are list-decodable from an appropriately large fraction of deletions and insertions with alphabet size dpeending only on these two fractions. Using this outer code reduces the problem at hand to showing the existence of arbitrarily large (though not necessarily positive-rate) binary codes with very good list decoding properties against deletions and insertions. 
This is established through a careful analysis of the list-decoding properties of the Bukh-Ma codes defined in~\eqref{eq:bukhma} and already discussed above in the context of Theorem~\ref{thm:thresholdlb}. We recall that plain Bukh-Ma inner codes were not enough to prove Theorem~\ref{thm:thresholdlb}, leading only to the (already non-trivial) lower bound $d^\star\geq 1/3$, although they turn out to be enough in the list-decoding setting.



\paragraph{The zero-rate threshold under relaxed adversarial models}

Guruswami and Li~\cite{GL18} considered zero-rate thresholds for error models between random deletions (under which positive rate communication with vanishing error probability is possible for any fraction of deletions $d<1$) and adversarial deletions.
Namely, they show that if one requires only vanishing average decoding error probability against adversarial deletions, then the zero-rate threshold is $1$, matching the result for random deletions. 
Here, ``vanishing average decoding error probability against adversarial deletions'' means that under any deletion pattern of up to $dn$ deletions, all but a vanishing fraction of codewords can be correctly recovered from the channel output. 
They also considered \emph{online} deletions, under which the adversary decides whether to delete or keep the $i$-th input bit $x_i$ (up to a budget of $dn$ deletions) based only on the values of $x_1,x_2,\dots,x_i$ (instead of having access to the whole input string as in the standard adversarial setting). 
They show that the zero-rate threshold for online deletions (under vanishing error probability) is closely related to the zero-rate threshold for adversarial deletions. In fact, one is below $1/2$ if and only if the other one is too.

\paragraph{The zero-rate threshold for sticky channels}
As the last topic in this section, we will discuss some results regarding the zero-error capacity of sticky channels. Unsurprisingly, since sticky errors appear to be easier to handle than deletions and insertions, much more is known about their zero-rate thresholds and zero-error capacities.
The zero-error capacity of channels with replication errors was considered first by Jain, Farnoud, Schwartz, and Bruck~\cite{JFSB17}. However, their main focus was on tandem duplications (where blocks of input bits, instead of single bits, are replicated) and on replication distributions with unbounded support over the positive integers. Sticky channels as considered in this survey correspond to channels that replicate blocks of size $1$ only. 

It is easy to see, as observed by Kova\v{c}evi\'c~\cite{Kov19}, that the zero-error capacity of sticky channels with replication distributions having full support over the positive integers is $0$. An adversary can insert replication errors to make all the runs of $0$'s and $1$'s have the same size across all the codewords of any candidate code. This means the only information we can gather from the channel output is the number of runs of $0$'s and $1$'s in the input codeword.
The case where the replication distribution $R$ satisfies $\Pr[R=i]>0$ if and only if $1\leq i\leq r$ for some integer $r$ is considerably more challenging. Kova\v{c}evi\'c~\cite{Kov19} was able to fully characterize the zero-error capacity of such channels.
\begin{thm}[\cite{Kov19}]\label{thm:zeroerrorsticky}
    The zero-error capacity of the sticky channel with replication distribution $R$ as above equals $\log(1/\rho)$, where $\rho$ is the unique real solution to the equation
    \begin{equation*}
        \sum_{j=1}^\infty x^{\frac{(r+1)^j-1}{r}}=1.
    \end{equation*}
\end{thm}
In contrast to Theorem~\ref{thm:zeroerrorsticky}, we recall that even the zero-rate threshold of adversarial binary deletions, which is an easier quantity to understand than the zero-error capacity, is still unknown. We note that this characterization can be extended to larger alphabets and tandem replications~\cite{Kov19}.

We conclude this section with a few significant related open problems:
\begin{itemize}
    \item Is the zero-rate threshold for adversarial deletions equal to $1/2$?
    
    \item What is the expected length of the LCS of two independent, uniformly random strings?
\end{itemize}

\bibliographystyle{IEEEtran}
\bibliography{survey-refs}

\end{document}